\documentclass{aa}  

\usepackage{graphicx}
\usepackage{txfonts}
\usepackage{multirow}
\usepackage[table,xcdraw]{xcolor}
\usepackage{booktabs}
\usepackage{xcolor}
\usepackage{upgreek}
\usepackage{longtable}

\usepackage[colorlinks=true, linkcolor=blue, citecolor=blue, urlcolor=blue]{hyperref}
\providecommand{\bjdtdb}{\ensuremath{\rm {BJD_{TDB}}}}
\providecommand{\tjdtdb}{\ensuremath{\rm {TJD_{TDB}}}}

\providecommand{\me}{\ensuremath{\,M_{\rm E}}}
\providecommand{\re}{\ensuremath{\,R_{\rm E}}}
\providecommand{\fave}{\langle F \rangle}
\providecommand{\fluxcgs}{10$^9$ erg s$^{-1}$ cm$^{-2}$}

\begin{document}

   \title{Hot Rocks Survey I: A possible shallow eclipse for LHS~1478~b}


\author{
P. C. August\inst{1} $^{\href{https://orcid.org/0000-0003-3829-8554}{\includegraphics[scale=0.5]{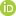}}}$,
L. A. Buchhave\inst{1} $^{\href{https://orcid.org/0000-0003-1605-5666}{\includegraphics[scale=0.5]{orcid.jpg}}}$,
H. Diamond-Lowe\inst{1,2} $^{\href{https://orcid.org/0000-0001-8274-6639}{\includegraphics[scale=0.5]{orcid.jpg}}}$,
J. M. Mendonça\inst{1,3,4} $^{\href{https://orcid.org/0000-0002-6907-4476}{\includegraphics[scale=0.5]{orcid.jpg}}}$,
A. Gressier\inst{2} $^{\href{https://orcid.org/0000-0003-0854-3002}{\includegraphics[scale=0.5]{orcid.jpg}}}$,
A. D. Rathcke\inst{1} $^{\href{https://orcid.org/0000-0002-4227-4953}{\includegraphics[scale=0.5]{orcid.jpg}}}$,
N. H. Allen\inst{5} $^{\href{https://orcid.org/0000-0002-0832-710X}{\includegraphics[scale=0.5]{orcid.jpg}}}$,
M. Fortune\inst{6} $^{\href{https://orcid.org/0000-0002-8938-9715}{\includegraphics[scale=0.5]{orcid.jpg}}}$,
K. D. Jones\inst{7} $^{\href{https://orcid.org/0000-0002-2316-6850}{\includegraphics[scale=0.5]{orcid.jpg}}}$,
E. A. Meier Vald\'es\inst{7} $^{\href{https://orcid.org/0000-0002-2160-8782}{\includegraphics[scale=0.5]{orcid.jpg}}}$, 
B.-O. Demory\inst{7,8} $^{\href{https://orcid.org/0000-0002-9355-5165}{\includegraphics[scale=0.5]{orcid.jpg}}}$,
N. Espinoza\inst{5, 2} $^{\href{https://orcid.org/0000-0001-9513-1449}{\includegraphics[scale=0.5]{orcid.jpg}}}$, 
C. E. Fisher\inst{9} $^{\href{https://orcid.org/0000-0003-0652-2902}{\includegraphics[scale=0.5]{orcid.jpg}}}$,
N. P. Gibson\inst{6} $^{\href{https://orcid.org/0000-0002-9308-2353}{\includegraphics[scale=0.5]{orcid.jpg}}}$,
K. Heng\inst{10, 11, 12, 13} , 
J. Hoeijmakers\inst{14} $^{\href{https://orcid.org/0000-0001-8981-6759}{\includegraphics[scale=0.5]{orcid.jpg}}}$,
M. J. Hooton\inst{15} $^{\href{https://orcid.org/0000-0003-0030-332X}{\includegraphics[scale=0.5]{orcid.jpg}}}$,
D. Kitzmann\inst{16} $^{\href{https://orcid.org/0000-0003-4269-3311}{\includegraphics[scale=0.5]{orcid.jpg}}}$,
B. Prinoth\inst{17} $^{\href{https://orcid.org/0000-0001-7216-4846}{\includegraphics[scale=0.5]{orcid.jpg}}}$,
J. D. Eastman\inst{18} $^{\href{https://orcid.org/0000-0003-3773-5142}{\includegraphics[scale=0.5]{orcid.jpg}}}$,
R. Barnes\inst{19} $^{\href{https://orcid.org/0000-0001-6487-5445}{\includegraphics[scale=0.5]{orcid.jpg}}}$
}

  \institute{
\label{inst:1} Department of Space Research and Technology, Technical University of Denmark, Elektrovej 328, 2800 Kgs.\,Lyngby, DK \and
\label{inst:2} Space Telescope Science Institute, 3700 San Martin Drive, Baltimore, MD 21218, USA \and
\label{inst:3} Department of Physics and Astronomy, University of Southampton, Highfield, Southampton SO17 1BJ, UK \and
\label{inst:4} School of Ocean and Earth Science, University of Southampton, Southampton, SO14 3ZH, UK \and
\label{inst:5} Department of Physics and Astronomy, Johns Hopkins University, 3400 N. Charles Street, Baltimore, MD 21218, USA \and
\label{inst:6} School of Physics, Trinity College Dublin, University of Dublin, Dublin 2, Ireland \and
\label{inst:7} Center for Space and Habitability, University of Bern, Gesellschaftsstrasse 6, 3012 Bern, Switzerland \and
\label{inst:8} Physikalisches Institut, University of Bern, Sidlerstrasse 5, 3012
Bern, Switzerland \and
\label{inst:9} Department of Physics, University of Oxford, Keble Road, Oxford, OX1 3RH, UK \and
\label{inst:10} Ludwig Maximilian University, Faculty of Physics, Scheinerstr. 1, Munich D-81679, Germany \and
\label{inst:11} ARTORG Center for Biomedical Engineering Research, University of Bern, Murtenstrasse 50, CH-3008, Bern, Switzerland \and
\label{inst:12} University College London, Department of Physics \& Astronomy, Gower St, London, WC1E 6BT, United Kingdom \and
\label{inst:13} University of Warwick, Department of Physics, Astronomy \& Astrophysics Group, Coventry CV4 7AL, United Kingdom \and
\label{inst:14} Lund Observatory, Division of Astrophysics, Department of Physics, Lund University, Box 118, 221 00 Lund, Sweden \and
\label{inst:15} Cavendish Laboratory, JJ Thomson Avenue, Cambridge CB3 0HE, UK \and
\label{inst:16} Space Research and Planetary Sciences, Physics Institute, University of Bern, Gesellschaftsstrasse 6, 3012 Bern, Switzerland \and 
\label{inst:17} Lund Observatory, Division of Astrophysics, Department of Physics, Lund University, Box 118, 221 00 Lund, Sweden \and
\label{inst:18} Center for Astrophysics | Harvard $\&$ Smithsonian, 60 Garden St, Cambridge, MA 02138, USA \and
\label{inst:19} Department of Astronomy, University of Washington, Seattle, WA 98105, USA
}

\authorrunning{P. C. August et al.}
   \date{Received XXXX; accepted YYYY}

 
  \abstract
   {M dwarf systems offer an opportunity to study terrestrial exoplanetary atmospheres due to their small size and cool temperatures. However, the extreme conditions imposed by these host stars raise question about whether their close-in rocky planets are able to retain any atmosphere at all.}
   {The Hot Rocks Survey aims to answer this question by targeting nine different M dwarf rocky planets spanning a range of planetary and stellar properties. Of these, LHS~1478~b orbits an M3-type star, has an equilibrium temperature of T$_{\rm eq} = 585\thinspace$K and receives $21$ times Earth's instellation.}
   {We observe two secondary eclipses of LHS~1478~b using photometric imaging at $15\upmu$m using the Mid-Infrared Instrument on the James Webb Space Telescope  (JWST MIRI) to measure thermal emission from the dayside of the planet. We compare these values to atmospheric models to evaluate potential heat transport and CO$_2$ absorption signatures.}
   {We find that a secondary eclipse depth of $138\pm 53\thinspace$ppm at the expected time for a circular orbit is preferred over a null model at $2.8\upsigma$, a moderate detection, though dynamical models do favour a non-eccentric orbit for this planet. The second observation results in a non-detection due to significantly larger unexplained systematics. Based on the first observation alone, we can reject the null hypothesis of the dark (zero Bond albedo) no atmosphere bare rock model with a confidence level of $3.3\upsigma$, though for $A_{\rm B}=0.2$ the significance decreases to $2.1\upsigma$. The tentative secondary eclipse depth is consistent with the majority of atmospheric scenarios we considered, spanning CO$_2$-rich atmospheres with surface pressures from $0.1$ to $10$ bar. However, we stress that the two observations from our program do not yield consistent results, and more observations are needed to verify our findings. The Hot Rocks Survey serves as a relevant primer for future endeavours such as the Director's Discretionary Time (DDT) Rocky Worlds program.}
   {}

   \maketitle
%
\section{Introduction}
The past decades of exoplanet hunting have shown that small Earth-sized planets are ubiquitous \citep{Fressin2013, Dressing2015b}. This is also true around M dwarfs, which are particularly well suited for studying rocky planets: not only are they very common in our solar neighbourhood \citep{Henry2018}, but their small size ($0.1$ to $0.6\thinspace$R$_{\odot}$) and lower effective temperatures ($2400$ to $3800\thinspace$K) make them favourable targets to observe transiting planets. 

The drawback of M dwarfs as planetary hosts is that they may not be as favourable for sustaining the atmospheres of rocky planets \citep{Shields2016}. Due to their lower masses, M dwarfs spend a prolonged period in the pre-main sequence (PMS) phase, a highly active and luminous stage. The PMS phase, followed by an extended X-ray saturation period, can significantly erode atmospheres accreted from the protoplanetary disk \citep[e.g., ][]{Krissansen-Totton2023}. These intense phases can strip away atmospheric components, particularly through processes such as hydrodynamic escape driven by stellar XUV flux. Flares, which are frequent in young and fast-rotating M dwarfs, can increase X-ray and UV (XUV) emissions by several orders of magnitude, leading to further heating and ionisation of the upper atmosphere and subsequent atmospheric loss \citep{Hawley2014, Medina2020, Medina2022b, doAmaral2022}. Additionally, stellar winds from M dwarfs are particularly harmful to planets within their habitable zones, which are much closer in than for solar-type stars \citep{Kopparapu2013}. While this is an observational advantage, it also means that these close-in planets face increased exposure to stellar winds, potentially causing additional atmospheric erosion \citep{Garcia-Sage2017, Dong2018}.

However, outgassing mechanisms might replenish and further shape the atmospheric composition of rocky planets \citep{Dorn2018, Herbort2020, Thompson2021, Tian_2023}. Planetary evolution and atmospheric chemistry models, alongside observations of rocky planets within our Solar System (e.g., Venus, Mars), suggest that CO$_2$ is likely to be a major constituent of these secondary atmospheres \citep{Gaillard2014, Krissansen-Totton2023, Tomberg2024}. In some cases, atmospheric erosion processes, such as the loss of lighter hydrogen atoms through hydrodynamic escape, can result in a significant O$_2$ build-up \citep{Luger2015b}. While this oxygen could indicate potential habitability, it might also be a signature of severe atmospheric loss rather than biological processes, especially on M dwarf planets where strong stellar activity drives atmospheric stripping.

Transit and eclipse spectroscopy, along with photometry, are the preferred observational methods to answer the question of whether or not M dwarf orbiting rocky planets can retain an atmosphere. However, transmission spectra derived from primary transits are vulnerable to M dwarf stellar activity which can affect the transmission spectra \citep[e.g.,][]{Lustig-Yaeger2023, Moran2023, Radica2024}. M dwarfs are cool enough that water can form in the photosphere, ultimately contaminating the planetary spectrum by mimicking atmospheric features \citep{Rackham2017, Rackham2018, Lim2023}. Many of the stellar contamination factors from M dwarf hosts are mitigated when moving into the infrared and observing secondary eclipses since what is measured is the pure flux contrast between the in- and out-of-eclipse configurations, thus avoiding issues related to stellar surface inhomogeneities. Additionally, stellar activity like flares have less of a significant impact at mid-infrared wavelengths than at optical or near-infrared, as the contrast between the hotter flaring regions and the cooler stellar photosphere diminishes in the Rayleigh-Jeans tail. On the downside, this method requires precise system parameter estimates, in particular the orbital and stellar properties. Only then can eclipse depths be converted into a dayside brightness temperature. The physical interpretation of this temperature in turn depends on atmospheric composition, heat redistribution and surface Bond albedo \citep{Cowan2011, Koll2019, Koll2022}. 

The question of rocky planet atmospheres around M dwarf stars motivated the need for a larger survey targeting rocky planets across a broad range of parameter space to test the hypothesis regarding the presence of atmospheres around these planets. The Hot Rocks Survey (PI Diamond-Lowe, Co-PI Mendonça, JWST GO 3730) focuses on a sample of nine different exoplanets observed in eclipse photometry at 15$\upmu$m with JWST/MIRI \citep{Redfield2024}. 

The ability of thermal emission measurements of M dwarf terrestrial planets to distinguish between thick atmosphere and likely bare rock scenarios was demonstrated with Spitzer \citep{Kreidberg2019, Crossfield2022, Zieba2022}. Since the launch of the James Webb Space Telescope, the MIRI instrument has returned a number of deep eclipse measurements, suggesting bare rock scenarios for multiple planets both in photometry \citep{Zieba2023, Greene2023}, and spectroscopy \citep{Mansfield2024, Xue2024, Zhang2024}. 

Transit spectroscopy attempts have also typically returned featureless spectra of rocky planets transiting M dwarfs, \citep[e.g.,][]{Lustig-Yaeger2023, Lim2023}, with the exception of the super-earth L98-59~d ($R_p = 1.58\thinspace R_{\oplus}$, $M_p = 2.31\thinspace M_{\oplus}$), and its neighbour, sub-earth L98-59~b ($R_p = 0.85\thinspace R_{\oplus}$), both showing evidence of a sulphur absorption feature \citep{Gressier2024, BelloArufe2025}. Recent observations of 55 Cancri~e ($R_p = 1.95\thinspace R_{\oplus}$, $M_p = 8.8\thinspace M_{\oplus}$) suggest the planet may have outgassed a secondary atmosphere composed of evaporated rock \citep{Hu2024, Patel2024}.

In this study, we focus on LHS~1478~b \citep{Soto2021}, an $R_p = 1.24\thinspace R_{\oplus}$, $M_p = 2.33\thinspace M_{\oplus}$ rocky exoplanet orbiting an M3-type star at a distance of $a = 0.018\thinspace$AU. It lies roughly in the middle of the Hot Rocks Survey sample in almost all considered parameters: equilibrium temperature, size, mass, proximity to the host star, and instellation. LHS~1478~b receives about $~20$ times the insolation of the Earth, and there are no additional planets known in the system. No major flaring activity or stellar rotational signature was observed in the TESS data. This is in stark contrast to the TRAPPIST-1 system where flaring activity over $10^{30}\thinspace$erg in the TESS bandpass has been estimated to $3.6$ flares per day \citep{Howard2023}. However, it is important to note that M dwarfs can appear inactive at optical wavelengths and still demonstrate flaring in the UV \citep[e.g.,][]{Loyd2018, Jackman2024}.

This paper is organised into eight sections. Section \ref{sec:observations} will briefly describe the observational details, while Sections \ref{sec:dataprocessing} and \ref{sec:datanalysis} are focused on the data reduction (from raw data to 1D time series) and light curve fitting process respectively. Section \ref{sec:modeling} addresses the modelling carried out to interpret the data, and we do a complete reanalysis of the system parameters by doing a joint fit with radial velocity and transit datasets in Section \ref{sec:systemparameters}. Section \ref{sec:results} goes through the main results. We discuss the limitations of our interpretations in Section \ref{sec:discussion} and conclude in Section \ref{sec:conclusions}.

\section{Observations}\label{sec:observations}
LHS-1478 b was observed as part of the JWST GO programme 3730 first on November 18, 2023, from 20:09:03 to 23:21:32 UTC and again on January 20, 2024, from 05:24:42 to 08:37:11 UTC. The observations used JWST/MIRI in time-series imaging mode with the F1500W filter, and were carried out using the sub256 subarray with the FASTR1 readout mode with 39 groups per integration resulting in a total of 964 integrations.

A baseline of 42 minutes, corresponding to the eclipse duration, was built in on either side of the time of secondary eclipse mid-point. An extra hour was added to the start of the observations to account for the uncertainty around the starting time of the observation, in addition to the 30 minutes pre-pended to account for any potential detector settling effects \citep{Morrison2023}. Finally, $18$ min of extra padding were added to the out-of-eclipse baseline to account for eccentricities up to 0.1 with $>90\%$ confidence. In total, this amounts to 192 minutes of observation time. LHS-1478 b does not have any known sibling planets that had to be factored into the phase constraints.

The eclipse depth, which had not been measured previously, was estimated in the Hot Rocks Survey proposal to be around $f = 300\thinspace$ppm for the dark bare rock case (i.e. zero albedo, no atmosphere), while simple atmospheric scenarios generated with \texttt{HELIOS} \citep{Malik2017, Malik2019a, Malik2019b} were predicted to give signals as deep as $150\thinspace$ppm for CO$_2$-based atmospheres. In order to differentiate between a bare rock and the presence of an atmosphere scenario at $3\upsigma$, two measurements were deemed necessary, reducing the error on the eclipse depth from around $60$ to $40\thinspace$ppm based on calculations performed with \texttt{Pandeia-2.0} \citep{Pontoppidan2016}.

\section{Data processing}\label{sec:dataprocessing}
We reduced the data with three different pipelines (\texttt{Eureka!}, \texttt{Frida}, and \texttt{transitspectroscopy}) to validate the results and ensure robust extraction of the light curves and eclipse depths, given JWST's new instrumentation, evolving understanding of detector behaviour, and the relative novelty of data reduction pipelines for reducing MIRI Imaging time series. The community-standard pipeline, \texttt{Eureka!}, is in active development and designed to be broadly applicable across diverse observations, making it valuable to compare its results with those from other methods. The last reduction was performed independently to further mitigate bias in the analysis.

\subsection{Frida}

\begin{figure*}
    \centering
    \includegraphics[width=0.8\textwidth,clip]{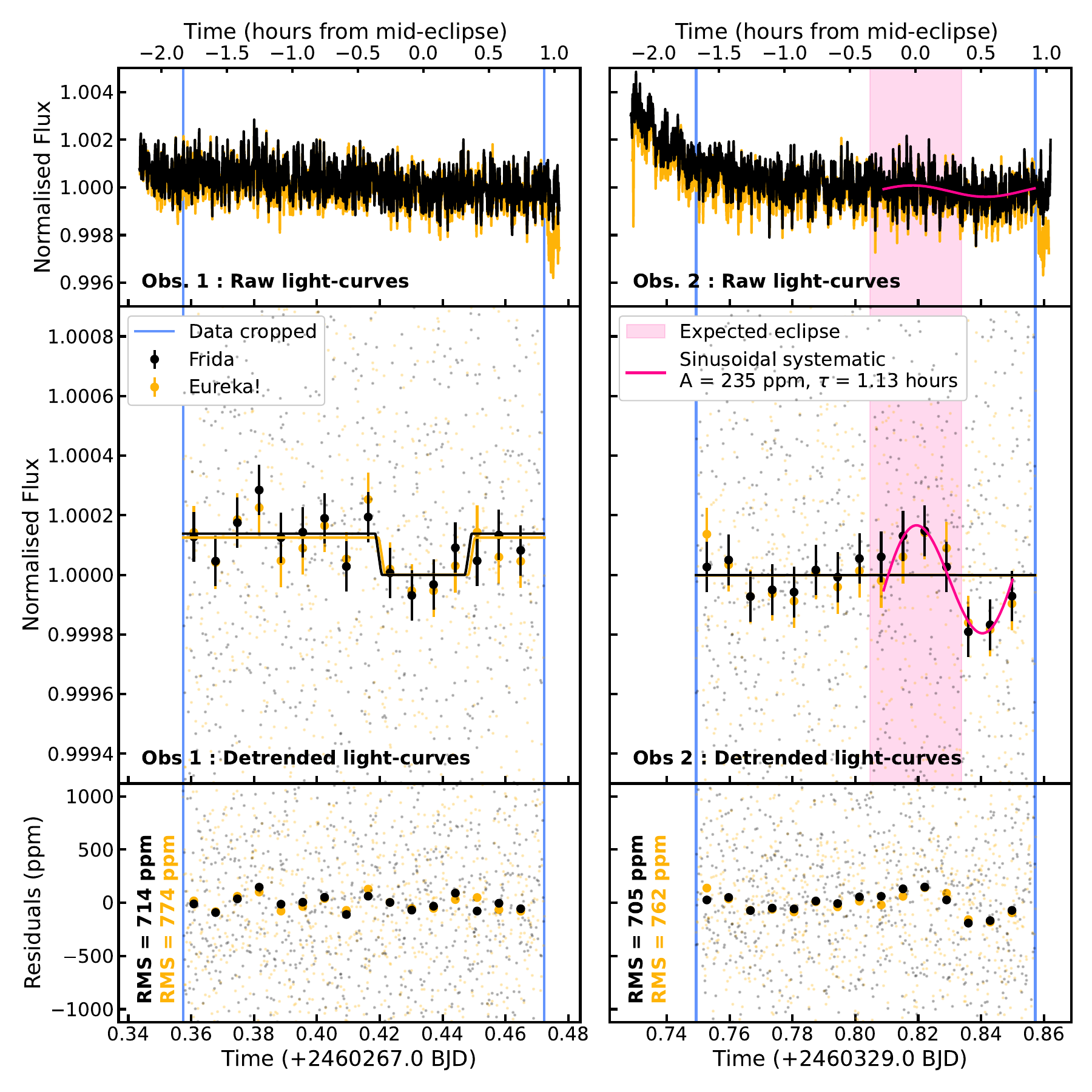}
    \caption{Raw and detrended light curves and residuals for the JWST/MIRI eclipse observations at $15\upmu$m for LHS-1478~b. The data and model from our nominal \texttt{Frida} reduction are shown in black, while the \texttt{Eureka!} equivalent is shown in orange. Note that in this plot the second visit was detrended using an exponential and a linear slope (EL) for \texttt{Frida}, and a simple linear slope (L) for \texttt{Eureka!}, after the detector settling slope was removed, and without an eclipse model. For the first visit we use a simple linear slope to model the systematics for both reductions, after discarding some initial and final integrations (see Section \ref{sec:datanalysis}). For the second visit, the expected eclipse timing, computed based on the first visit, is represented as a purple shaded region between ingress and egress, and we highlight in red the unexplained systematic (referred to later on as "sinusoidal-like" systematic).}
    \label{fig:lightcurves}
\end{figure*}
\texttt{Frida} is an end-to-end JWST pipeline developed for exoplanet transmission and emission spectroscopy and photometry. Stage 1 of the pipeline utilises the official JWST pipeline\texttt{jwst}\footnote{https://github.com/spacetelescope/jwst} for early-stage functionalities such as ramp fitting and the flat fielding, while the rest of \texttt{Frida} is custom-developed. 

Starting from stage 2, \texttt{Frida} does not utilise any routines from \texttt{jwst}. Cosmic rays are identified and removed using time-series analysis to detect 5-sigma outliers at the pixel-level light curves. They are then replaced with values smoothed by a Gaussian filter with a length of 10 integrations. Finally, we perform optimal extraction of the photometric images by using a normalised smoothed median-weighted profile (representing the point spread function) to define pixel weights. Although JWST has very stable pointing, a small oscillating drift is observed both along columns and rows with amplitudes of a few thousandths of a pixel. The normalised smoothed point spread function (PSF) is shifted accordingly at each integration to compensate for this drift.  

\texttt{Frida} can perform both classic aperture photometry as well as optimal extraction, which is implemented on a $20\thinspace$x$\thinspace20$ pixel square around the target and using a "z-cut", that is a flux level under which the pixels are discarded. The MIRI PSF is an intricate pattern roughly composed of an inner circular region, a secondary petal-like ring, and higher order patterns resembling a snowflake. The z-cut method allows us to capture the complex shape of the snowflake pattern without including background pixels. Both the square size and z-cut level are adjustable, and the z-cut is defined as a multiplier of the maximal flux value. After testing hexagonal and circular aperture photometry as well as optimal extraction with different z-cut levels, we found the latter to perform much better. Using the root mean square (RMS) and median absolute deviation (MAD) values to guide our choice of z-cut, we settled on $z_{cut} = 0.007$ for the first observation and $z_{cut} = 0.005$ for the second. These values capture both the primary and secondary rings (i.e. central circle and "petals") of the PSF. The final light curves are shown in Figure \ref{fig:lightcurves}, and the RMS and MAD values are reported in Table \ref{tab:datared}, along with the other pipelines.

\subsection{Eureka!}
We used \texttt{Eureka!-v0.10} \citep{Bell2022}, an end-to-end pipeline for time series observations (TSO) performed with JWST, to reduce our data, starting from the raw uncalibrated (uncal) files available on the Mikulski Archive for Space Telescopes. This open-source package is widely used for JWST observations as a whole \citep[e.g.,][]{August2023, Bean2023} as well as specifically for similar MIRI Imaging time series observations \citep{Zieba2023, Greene2023}.
The early stages of \texttt{Eureka!} are a wrapper for the \texttt{jwst} pipeline \citep{bushouse_2022_7229890} and so the stage 1 and stage 2 options were set according to the official JWST TSO guidelines. In stage 1, we set the \texttt{skip-firstframe} and \texttt{skip-lastframe} step to False, meaning that we include the first and last group in the ramp fitting of a given exposure, as it improved the flagging and removal of the cosmic rays. However, this introduced an offset between the last segment of the observations, consisting of only $36$ frames, and the rest of the photometric light curve (see Figure \ref{fig:lightcurves}). We later chose to remove this segment when performing fits to the light curve. For the jump step, we set the detection threshold at $5\upsigma$. 

The flux was extracted using aperture photometry (stage 3). After testing out different aperture radii (in pixels) to minimise the MAD of the resulting light curve we decided to settle for a $(\rm r_{\rm aper}, \rm r_{\rm bkg}^{\rm in}, \rm r_{\rm bkg}^{\rm out})$ combination of $(5, 20, 45)$ for the first visit and $(5, 20, 40)$ for the second. These configurations are set so that the aperture encompasses the inner part of the PSF, where most of the flux is concentrated, and the background annulus avoids most of the "snowflake"-shaped diffraction pattern. \texttt{Eureka!} also performs a double-iteration sigma clipping to remove outliers from the light curve. The thresholds for these are set at [6,6] and [5,5] for each observation, respectively, with little variations in the MAD around those values.

\subsection{transitspectroscopy}\label{subsec:transitspectroscopy}
We conducted an independent reduction of the two eclipse observations of LHS~1478~b, following the procedures outlined in Gressier et al. (in prep.) for the reduction of four eclipse observations of the Hot Rocks Survey target L231-32~b. This reduction used the open-source Python package \texttt{transitspectroscopy} \citep{espinoza_2022} with custom processing in Stage 2. In Stage 1, standard corrections were applied, skipping the reference pixel step for MIRI subarrays. A custom jump correction was implemented by identifying outliers through the subtraction of a median filter from the group differences.

In the custom Stage 2, raw integration images were converted into time series light curves. To estimate the background, the target signal was masked, and a median frame was computed. A median filter and a Gaussian filter were then applied to remove large-scale background structures and assess the background level. This smoothed background was subtracted from each frame. Outliers in the pixel light curves were flagged using a median filter and replaced with median values. A 2D Gaussian model was fitted to determine the centroid position, after which optimal aperture photometry was performed using a 10-pixel-radius circular aperture. The total flux within the aperture was calculated, uncertainties were propagated, and two light curves were obtained and normalised based on the final 100 integrations. This reduction serves as a baseline for the joint fit presented in \ref{subsec:jointfit}.

\begin{table}[]
\centering
\caption{RMS and MAD values of the raw light curves, in ppm.}
\begin{tabular}{@{}ccccccc@{}}
\toprule
        & \multicolumn{2}{c}{\texttt{Frida}} & \multicolumn{2}{c}{\texttt{Eureka!}} & \multicolumn{2}{c}{\texttt{transitsp.}} \\ 
        & \textbf{RMS}              & \textbf{MAD}             & \textbf{RMS}               & \textbf{MAD}              & \textbf{RMS}                     & \textbf{MAD}                    \\ \midrule
\textbf{Visit 1} & 820              & 531             & 813               & 569              & 975                     & 582                    \\
\textbf{Visit 2} & 1158              & 610             & 1022               & 620              & 2041                     & 784                    \\ \bottomrule
\end{tabular}\newline
\tablefoot{The RMS and MAD values are calculated on the light curves after removing the last 36 integrations}.
\label{tab:datared}
\end{table}

\section{Data analysis}\label{sec:datanalysis}
\subsection{Frida}\label{subsec:frida}
We perform a thorough analysis of the light curve to probe the robustness of the eclipse depth measurements with respect to the models used to treat the systematics. The general equation describing the full light curve model in \texttt{Frida} is given by Eq. \eqref{eq:frida_model}:
\begin{equation}\label{eq:frida_model}
    F_{lc}(t) = F_{lc}^{\texttt{batman}}(t) + F_{sys}(t),
\end{equation}
where $F_{lc}^{\texttt{batman}}(t)$ represents the light curve, parametrised using the \texttt{batman} package \citep{Kreidberg2015}, and $F_{sys}(t)$ is the model for the systematics. A range of models are available in \texttt{Frida}, from simple polynomial models to exponential slopes and GPs, which can also all be combined. There is also the possibility for the user to write up their own desired model for the systematics.

For the first observation, we tried a first and second order polynomial, a linear slope with an exponential (motivated by the detector settling slope), and a GP with a squared exponential kernel using the \texttt{george} package \citep{Ambikasaran2015}. Other models, including more complex sum of exponentials and higher order polynomials were considered during the analysis, but discarded due to poorer BIC performances.

We use nested sampling with \texttt{UltraNest}\footnote{\url{https://johannesbuchner.github.io/UltraNest/}} \citep{Buchner2021} to fit the time series observation. We first used large, uniform priors on the eclipse depth $f_p/f_*$, also allowing for negative eclipse depths (from $-100$ to $500$~ppm), and time of secondary eclipse $t_{sec}$ (full observation range). As a second step, once the eclipse was identified, we used a uniform prior with half an eclipse duration on either side of the expected secondary eclipse time, and positive eclipse depths going up to $300$~ppm (hot bare rock case). In our final run we use a linear slope for the systematics, fix all orbital parameters to the values obtained through the joint \texttt{Exofastv2} fit (see Section \ref{sec:systemparameters}), and we fit for the eclipse depth $f_p/f_*$ and the time of secondary eclipse $t_{sec}$ (see Table \ref{tab:fridafit} for a summary of the corresponding priors and posteriors).

While an extra $30$~minutes is built into the observation to account for settling, the exact behaviour of this effect is not well understood. In particular, observations have shown significant variations in the strength, duration, and even sign of these "exponential ramps" \citep[e.g.,][]{Zieba2023, Greene2023}. For this reason we explore the effect of removing parts of the initial data affected by the detector ramp effect (clipping), by removing $n_{clip} = 100$ to $200$ integrations (corresponding roughly to $t_{clip} = 20$ to $40$~minutes) at the beginning of the observation and running our fitting routine again. For the first observation, $n_{clip} = 200$ integrations remove the ramp entirely and thus don't call for models more complex than the linear case.

For the first visit, we removed the first $100$ and last $36$ integrations of the light curve (last segment) and then fit a linear slope to the data. This particular configuration is a result of various tests comparing the RMS and MAD values of the light curve as well as inter-model comparison using the Bayesian information criterion (BIC) and reduced chi-square statistics (see Table \ref{tab:data_reduction}). We detect a possible eclipse with a depth of $f_p/f_* = 138 \pm 53$~ppm. This model is preferred over a simple linear slope with $\Delta\log Z = 2.41$. Using the $\chi^2$-distribution, we find this eclipse model is preferred over a simple mean of the detrended data at $2.8~\sigma$. For the same clipping, a GP performs equally well in terms of $\chi^2_{red}$ , BIC, and RMS and finds an eclipse depth of $f_p/f_* = 111 \pm 65$~ppm. All the methods we tested found the eclipse depth within $<1\upsigma$ of each other, as can be seen in Table \ref{tab:data_reduction}. The inter-model comparison diagnostics are also nearly identical.

For the second visit, no amount of GPs, pre- or post-processing allowed us to mitigate the correlated noise in the light curve. The eclipse remains undetectable, and the residuals show some sinusoidal-like behaviour reaching a peak exactly around the time where the eclipse should be. In fact, allowing for negative $f_p/f_*$ values in the fits typically gave fully converged negative eclipse values. The dip of the sinusoidal signal right at the point of predicted egress mimics an eclipse signal, which the models favour if the time of secondary eclipse is left as a free parameter. However, we stress that this signal is likely not an eclipse, but rather a strong systematic.

\begin{table}[]
\centering
\caption{Priors and posteriors of the \texttt{Frida} light curve fit for the first visit.}
\setlength{\tabcolsep}{5pt}
\begin{tabular}{@{}lll@{}}
\toprule
{\textbf{Parameter}}    & {\textbf{Uniform prior}} & { \textbf{Posterior}}                     \\ \midrule
{ $f_p/f_*~[\rm ppm]$}                 & { [0,300]}       & { $138^{+52}_{-55}$}                      \\
{ $t_{sec}~[- 2460267.0~\rm BJD]$} & { [0.417,0.447]} & { $0.434^{+0.006}_{-0.005}$}              \\
{ $c_0$}                 & { [-0.1,0.1]}    & { $(5.64^{+0.69}_{-0.66})\cdot 10^{-4}$}  \\
{ $c_1$}                 & { [-1,1]}        & { $(-7.85^{+0.71}_{-0.69})\cdot 10^{-3}$} \\ \bottomrule
\end{tabular}\newline
\tablefoot{Posteriors represented as the median and $1\upsigma$ uncertainties. $c_0$ and $c_1$ are the intercept and slope coefficients for the linear systematic model.}
\label{tab:fridafit}
\end{table}

\subsection{Eureka!}
We also use nested sampling via \texttt{dynesty} in the \texttt{Eureka!} fitting routine. Instead of summing the \texttt{batman} light curve model with the systematics as in Equation \eqref{eq:frida_model}, \texttt{Eureka!} multiplies these functions. The end result is effectively the same, but the absolute values of the systematics-related parameters are not directly comparable.

Here again, we first let the secondary eclipse time free across the whole time series and allow for negative eclipse depths, and then refine the priors similarly to what was described in Section \ref{subsec:frida}. \texttt{Eureka!} records the positions and deviations of the centroid in $x$ and $y$ and allows us to decorrelate the light curve against them using multiplicative coefficients when fitting the light curve. We use a linear model both with and without including these extra parameters, as well as a GP. The default GP in \texttt{Eureka!} is a `Matern32' kernel with \texttt{celerite} \citep{celerite}. We also remove $n_{clip} = 100, 150$ and $200$ integrations to ensure robustness and comparability across pipelines.

For the first visit, we again chose to remove the first $100$ and last $36$ integrations and preferred the simpler linear trend over the other models. This gives us an eclipse depth of $f_p/f_* = 125\pm 56$~ppm, which is in agreement with the \texttt{Frida} result within the uncertainties (see comparative corner plot in Figure~\ref{fig:cornerplot}). In this case, the GP overall agrees with the linear trend. Including the positional correlators gave overall slightly lower eclipse depths (on the order of $110$~ppm), but were still statistically equivalent to the other results. It is important to note that the linear with centroid decorrelation (LPOS) models had significantly higher BIC values as they introduce four additional parameters to the fit compared to the simple linear (L) model (see Table \ref{tab:data_reduction}).

The second observation exhibits the same strong systematics with the \texttt{Eureka!} data reduction. Consequently, the fitting routine was unable to detect an eclipse, regardless of the combination of models and clipping tested out for this visit.

\subsection{Joint fit with juliet}\label{subsec:jointfit}
A joint fit of the two light curves reduced with \texttt{transitspectroscopy} as described in Section \ref{subsec:transitspectroscopy} was performed using the Python package \texttt{juliet} \citep{Espinoza2019b} incorporating \texttt{batman} \citep{Kreidberg2015} for transit and eclipse modelling, and \texttt{dynesty} \citep{Speagle2020} for nested sampling. Orbital parameters were fixed to the values from \citet{Soto2021}. We excluded the first 150 and the last 36 integrations, applying a range of different models : an exponential and linear detrending model (EL), a GP with a Matern32 kernel and a linear detrending model (LGP), and a GP with a Matern32 kernel and an exponential detrending model (EGP). The results are shown in Table~\ref{tab:data_reduction}. A joint fit for the eclipse depth was conducted, using separate detrending models per observation, yielding an eclipse depth of $f_p/f_* = 69^{+59}_{-45}~$ppm, which corresponds to the average between the first eclipse and zero.
\begin{table}[]
\centering
\caption{Inter-model comparison for the light curve fitting.}
\label{tab:data_reduction}
\setlength{\tabcolsep}{5pt}
\begin{tabular}{lclcccc}
\toprule
\textbf{Pipeline} & \textbf{$n_{clip}$} & \textbf{Model}   & \textbf{BIC}      &\textbf{$\chi_{red}^2$} &\textbf{RMS}& \textbf{$f_p/f_*$}   \\ \midrule
\multirow{3}{*}{\texttt{Frida}} & 100 & L & \cellcolor[HTML]{FFC8E4}-9561&\cellcolor[HTML]{FFC8E4}1.44&\cellcolor[HTML]{FFC8E4}714&\cellcolor[HTML]{FFC8E4}$138\pm 53$ \\
                                &      & EL &-9554&1.44&714&$128\pm 56$    \\
                                &      & GP  &-9546&1.43 &713&$110\pm 63$   \\ \cmidrule(l){2-7}
                                & 150  & L   & -8984&1.44&713&$138\pm 52$    \\
                                &      & EL  &-8977&1.44&714&$124\pm 57$    \\
                                &      & GP  & -8969 &1.43&713&$111\pm 65$    \\ \cmidrule(l){2-7}
                                & 200  & L   & -8400&1.45&715&$139\pm 53$    \\ \midrule
\multirow{3}{*}{\texttt{Eureka!}} & 100  & L   &-9442&1.51&774&$125\pm 56$    \\
                                  &      & LPOS &-9303&1.47&762&$112\pm 59$    \\
                                  &      & GP  &-9325&1.64&806&$110\pm 67$    \\ \cmidrule(l){2-7}
                                  & 150  & L   &-8877&1.50& 772&$124\pm 56$    \\
                                  &      & LPOS &-8891&1.46&759&$107\pm 59$   \\
                                  &      & GP  &-8852&1.62&801&$109\pm 70$    \\ \cmidrule(l){2-7}
                                  & 200  & L   &-8307&1.51&772&$127\pm 60$    \\
                                  &      & LPOS &-8327&1.45&756&$107\pm 59$    \\ \midrule
\multirow{2}{*}{\texttt{juliet$^{*}$}} & 150  & EL   & -8880 & 1.35  & 784 & $101_{-59}^{+80}$   \\
&  & LGP   & -8880   & 1.34  &  782  & $107_{-61}^{+69}$    \\
&  & EGP   & -8874   & 1.34  &  782  & $106_{-61}^{+69}$    \\
\bottomrule
\end{tabular}\newline
\tablefoot{We fit the observation 1 light curve obtained through different data reduction pipelines with their respective fitting routines. The different functions explored to model the systematics are:\\
L, a first order polynomial; EL, a first order polynomial with an exponential term; GP, a gaussian process (squared exponential kernel for \texttt{Frida} and Matern32 kernel for \texttt{Eureka!}; LPOS, and a first order polynomial with centroid position decorrelation; LGP, a first order polynomial and a GP (Matern32 kernel); EGP, an exponential and a GP (Matern32 kernel). \\
The BIC is computed using BIC$= -2\cdot\log(L) + k\cdot\log(N)$, where $\log(L)$ is the maximum log-likelihood for each fit, $k$ is the number of free parameters and $N$ is the number of integrations in the light curve. The final $f_p/f_*$ value quoted in the paper and in Figure \ref{fig:atm_models} is highlighted in pink. \\
$n_{clip}$ are the integrations discarded at the start of the light curve. \\
\tablefoottext{*}{This fit was performed on the light curve reduced independently with the \texttt{transitspectroscopy} pipeline.}}
\end{table}
\subsection{Noise characterisation in the light curves}
The second visit has much higher RMS and MAD values (see Table \ref{tab:datared}), but this difference is mostly due to the significant detector settling ramp over the first $~30$ minutes of observation. The Allan deviation plots in Figure \ref{fig:allandeviation} show that, once the ramp is removed and the light curve is detrended, both observations exhibit similar scatter. Even a linear fit is sufficient to remove most of the long-term systematics in the second visit. In terms of white noise, the two observations are equivalent.

The Allan plots and RMS values are not good diagnostics of short-term, small amplitude, correlated noise. In order to understand the systematic masking the eclipse in the second visit, we cut out about $1.5$ hours of data and fit a sinusoid to the data points using \texttt{emcee} \citep{emcee}, allowing us to estimate a timescale and an amplitude for the systematic. We find that it is best described by a sine function with an amplitude of $A = 235~$ppm and a timescale of $\tau = 1.13~$hours. Both of these are of the same order of magnitude as the eclipse signal we are trying to recover. Additionally, the peak of the systematic occurs around the mid-eclipse time, and the drop happens at egress, going against the shape of the eclipse signal. Without knowledge of the exact nature of the systematic, we are not able to accurately detect and measure a planetary signal at this time in the light curve. The injection recovery tests presented in Section~\ref{subsec:injrecov} further support this statement.

\subsection{Injection recovery for the second observation}\label{subsec:injrecov}
We performed injection recovery tests using \texttt{Frida} to put an upper limit on the eclipse depth that we would be able to recover given the significant systematics affecting the light curve. Two series of injection recoveries were performed: one where we injected the signal at the expected time of eclipse based on visit one ($t_{\rm expected} = 2460329.817~$BJD), and another where we injected it in the most well behaved part of the light curve (i.e., not in the ramp, and not in the sinusoidal systematic, we choose $t_{\rm behaved} = 2460329.800~$BJD). The results are summarised in Table \ref{tab:injection_recoveries}.

When injecting the signal at $t_{\rm behaved}$ and fitting a linear model, we cropped the first and last $150$ integrations, to get rid of both the detector settling ramp and the late sinusoidal systematic. For the $t_{\rm expected}$ injection recoveries, we clip the first $150$ and last $36$ integrations as we do in the standard analysis (Section~\ref{subsec:frida}). For the GP, we remove the first $100$ and last $36$ integrations.

The tests show that an eclipse signal at the expected timing in the light curve is undetectable below $300\thinspace$ppm. Such a signal could eventually be recovered and correctly measured, albeit with large error bars, by fixing, or at the very least strongly constraining, the secondary eclipse time. This is easily explained by the specific behaviour of the correlated noise around these critical times of secondary eclipse and egress, which strongly affects the eclipse.
However, if injected into the "well-behaved" part of the light curve, even a $100\thinspace$ppm signal could eventually be recovered with the help of a GP as well as some prior knowledge regarding the timing of the eclipse. 
The main reason we cannot recover the eclipse signal in the second observation is due to a large unexplained systematic that occurs precisely around the time of the eclipse event. This also suggests that the detector settling behaviour noted in Figure \ref{fig:lightcurves} can be treated by removing the first part of the ramp and using first order polynomials and exponential terms to model out the remaining long term trend.

We also perform injection recovery tests on the first visit, at $t = 2460267.38~$BJD, where a smaller, but similar systematic occurs. We mask out the actual eclipse by injecting an equal and opposite eclipse signal into the light curve and fit for different systematic models and an eclipse. We find the injected eclipse is undetectable below $200~$ppm and poorly measured ($>1.5\upsigma$ discrepant) below $300~$ppm. In other words, had the eclipse occurred at the time of this systematic, it would have impeded our ability to detect it, similar to the situation in the second visit.

\begin{table}[]
\centering
\caption{Diagnostic table of the injection recovery tests.}
\begin{tabular}{llllcc}
\toprule
$f_{inj}$             & $t_{inj}$              & $t_{sec}$               & \textbf{Model} & $\sigma_{det.}$           & $\sigma_{meas.}$        \\ \midrule
                      &                        &                         & L   & \cellcolor[HTML]{FF6961}-   & \cellcolor[HTML]{FF6961}-      \\
                      &                        & \multirow{-2}{*}{free}  & GP  & \cellcolor[HTML]{FF6961}-   & \cellcolor[HTML]{FF6961}-      \\
                      &                        &                         & L   & \cellcolor[HTML]{FF6961}$1.3$   & \cellcolor[HTML]{FF6961}-      \\
                      & \multirow{-4}{*}{exp.} & \multirow{-2}{*}{fixed} & GP  & \cellcolor[HTML]{FF6961}$1.3$   & \cellcolor[HTML]{FF6961}-      \\ \cmidrule(l){2-6}
                      &                        &                         & L   & \cellcolor[HTML]{FF6961}-   & \cellcolor[HTML]{FF6961}-      \\
                      &                        & \multirow{-2}{*}{free}  & GP  & \cellcolor[HTML]{F8D66D}$1.5$  & \cellcolor[HTML]{8CD47E}$<1.0$   \\
                      &                        &                         & L   & \cellcolor[HTML]{8CD47E}$2.4$  & \cellcolor[HTML]{8CD47E}$<1.0$  \\
\multirow{-8}{*}{150} & \multirow{-4}{*}{beh.} & \multirow{-2}{*}{fixed} & GP  & \cellcolor[HTML]{F8D66D}$1.7$  & \cellcolor[HTML]{8CD47E}$<1.0$  \\ \midrule
                      &                        &                         & L   & \cellcolor[HTML]{FF6961}$<1.0$   & \cellcolor[HTML]{FF6961}-      \\
                      &                        & \multirow{-2}{*}{free}  & GP  & \cellcolor[HTML]{FF6961}$<1.0$   & \cellcolor[HTML]{FF6961}-      \\
                      &                        &                         & L   & \cellcolor[HTML]{FF6961}$<1.0$   & \cellcolor[HTML]{FF6961}-      \\
                      & \multirow{-4}{*}{exp.} & \multirow{-2}{*}{fixed} & GP  & \cellcolor[HTML]{FF6961}$<1.0$   & \cellcolor[HTML]{FF6961}-      \\ \cmidrule(l){2-6}
                      &                        &                         & L   & \cellcolor[HTML]{FF6961}$<1.0$   & \cellcolor[HTML]{FF6961}-      \\
                      &                        & \multirow{-2}{*}{free}  & GP  & \cellcolor[HTML]{F8D66D}$1.5$  & \cellcolor[HTML]{8CD47E}$<1.0$   \\
                      &                        &                         & L   & \cellcolor[HTML]{8CD47E}$3.3$   & \cellcolor[HTML]{8CD47E}$<1.0$  \\
\multirow{-8}{*}{200} & \multirow{-4}{*}{beh.} & \multirow{-2}{*}{fixed} & GP  & \cellcolor[HTML]{F8D66D}$2.0$    & \cellcolor[HTML]{8CD47E}$<1.0$  \\ \midrule
                      &                        &                         & L   & \cellcolor[HTML]{F8D66D}$1.5$  & \cellcolor[HTML]{FF6961}$>3.0$   \\
                      &                        & \multirow{-2}{*}{free}  & GP  & \cellcolor[HTML]{F8D66D}$1.5$  & \cellcolor[HTML]{F8D66D}$1.3$  \\
                      &                        &                         & L   & \cellcolor[HTML]{F8D66D}$2.0$    & \cellcolor[HTML]{FF6961
}$>3$   \\
                      & \multirow{-4}{*}{exp.} & \multirow{-2}{*}{fixed} & GP  & \cellcolor[HTML]{F8D66D}$1.8$  & \cellcolor[HTML]{F8D66D}$1.6$  \\ \cmidrule(l){2-6}
                      &                        &                         & L   & \cellcolor[HTML]{8CD47E}$5.2$  & \cellcolor[HTML]{8CD47E}$<1.0$  \\
                      &                        & \multirow{-2}{*}{free}  & GP  & \cellcolor[HTML]{F8D66D}$2.0$    & \cellcolor[HTML]{8CD47E}$<1.0$  \\
                      &                        &                         & L   & \cellcolor[HTML]{8CD47E}$5.3$  & \cellcolor[HTML]{8CD47E}$<1.0$  \\
\multirow{-8}{*}{300} & \multirow{-4}{*}{beh.} & \multirow{-2}{*}{fixed} & GP  & \cellcolor[HTML]{F8D66D}$2.4$  & \cellcolor[HTML]{8CD47E}$<1.0$  \\ \bottomrule
\end{tabular}\newline
\tablefoot{We injected eclipses of $f_{inj} = 150\thinspace$ppm (expected signal based on the first observation), $f_{inj} = 200\thinspace$ppm, and $f_{inj} = 300\thinspace$ppm (expected bare rock signal). The second column indicates whether the injection was done at the expected secondary eclipse time (exp.) or in a more well-behaved part of the light curve (beh.). The third column shows whether or not the prior on $t_{sec}$ was left open or if it was fixed to the time of injection $t_{inj}$, and the fourth column denotes the choice of the model when fitting the light curve. Finally, the last two columns show whether the eclipse was detected (how many standard deviations away from $0$), and whether it was well measured (how many standard deviations away from $f_{inj}$). The red-yellow-green colour code highlights the significance of the detection and measurement for visual clarity.}
\label{tab:injection_recoveries}
\end{table}

\section{Modelling}\label{sec:modeling}
\begin{figure*}
    \centering
    \includegraphics[width=\textwidth,clip]{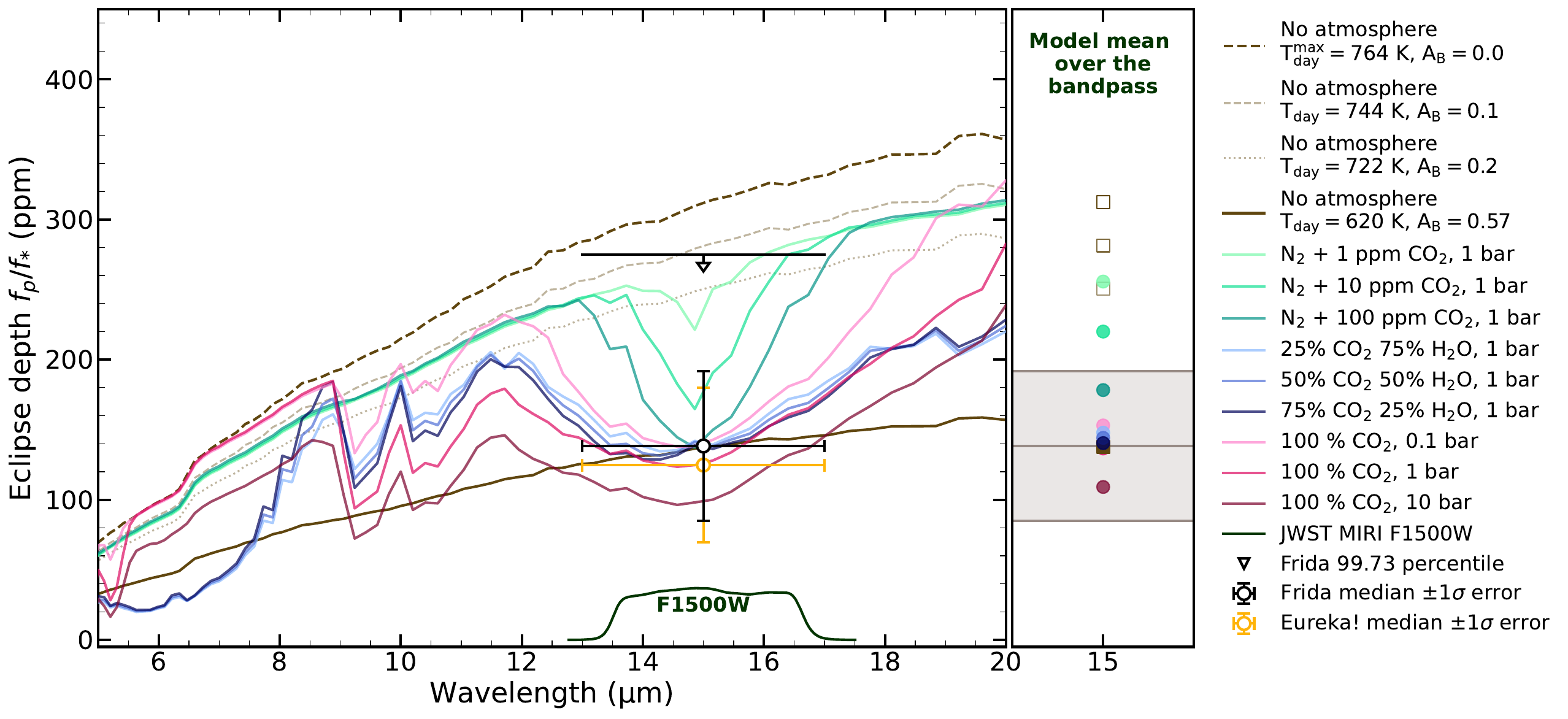}
    \caption{\textit{Left panel:} JWST/MIRI photometry eclipse depth measurement at 15$\thinspace\upmu$m (black data point) in context with different atmospheric models (coloured full lines), and airless blackbody (BB) models for different Bond albedos ($A_\mathrm{B}$; brown dashed and dotted lines). $T_{day}$ is the corresponding dayside brightness temperature as computed with Equation \eqref{eq:Tday}. All atmospheric scenarios are computed with $A_\mathrm{B} = 0.1$ and using the "f-approximation formula" based on \cite{Koll2022}. N$_2$ and O$_2$ are both spectrally inactive (no significant absorption features) at these wavelengths, thus yielding nearly identical thermal emission spectra. Only the N$_2$/CO$_2$ scenarios are plotted for simplicity. The result from the \texttt{Eureka!} reduction is also shown (orange data point). \textit{Right panel:} Atmospheric models (coloured circles) and atmosphere-less models (squares) from the left panel averaged over the F1500W bandpass. The gray shaded region represents the lower and upper error on the photometric measurement (solid line) obtained through a simple mean of both pipeline results.}
    \label{fig:atm_models}
\end{figure*}

Our atmospheric models are computed using \texttt{HELIOS} \citep{Malik2017, Malik2019a, Malik2019b}. The stellar spectrum used as an input for the models comes from an interpolation of the \texttt{SPHINX} model grid for M dwarfs \citep{Iyer2022, Iyer2023} to the temperature, $\log g$, and metallicity of the host star \citep{Soto2021}. 

We include CO$_2$, H$_2$O as the main molecular species in our models as well as N$_2$ as a neutral background gas. Their opacities are computed with the \texttt{HELIOS-K} module operating on the DACE opacity database \citep{Grimm2021}. The line lists for the gas absorption are HITEMP2010 for CO$_2$ \citep{Rothman2010}, BT2 for H$_2$O \citep{Barber2006}, and WCCRMT for N$_2$ \citep{Western2018}, and the Rayleigh scattering for the different molecules is based on \cite{Cox2000, Sneep2005, Thalman2014}.
The choice to focus on these molecules is rooted in atmosphere evolution theory and modelling predictions for M dwarf rocky planets. These planets are expected to have lost most, if not the entirety, of their H/He envelope during the long, high-energy pre-main sequence phase of their host star \citep{Owen2019}. Hydrogen, being so light, is highly susceptible to atmospheric escape, and may drag heavier elements along via hydrodynamic drag, resulting in significant atmospheric mass loss. Terrestrial planets may however develop secondary atmospheres through outgassing processes \citep{Tian_2023}. Depending on the initial formation of the mantle, the surface, the overall metallicity content, and the temperature, this would result in atmospheres dominated by CO$_2$ and H$_2$O \citep{Tomberg2024}, or N$_2$/O$_2$-dominated atmospheres with trace presence of CO$_2$ or H$_2$O \citep{Herbort2020}. 

We focused on CO$_2$ as a main heavy molecule constituent because it is predicted to be the dominant atmospheric molecule for highly irradiated rock planets \citep{Tian2009}. Other common, hydrogen-based molecules like water and methane are expected to dissociate in the upper atmosphere under the effect of incoming stellar radiation, leaving the oxygen and carbon atoms to recombine. CO$_2$ is a stable molecule, and also has the advantage of exhibiting a strong, detectable absorption feature at 15$\upmu$m \citep{Zasova2004, Mendonca2020}. The photometric bandpass covered by the JWST/MIRI F1500W filter specifically probes this feature. We assumed a Bond albedo of $A_\mathrm{B}=0.1$ for all of our atmospheric models, and used the "f approximation formula" based on \cite{Koll2022} built into \texttt{HELIOS}, running each model 4 times to have a converged solution for the optical depth, as recommended by the documentation. Additional models where we explore the impact of the extreme values of heat transport in the atmosphere are shown in the Appendix (see Figure \ref{fig:edgecasesf}).

For the no-atmosphere models, we follow the energy balance approach detailed in \cite{Malik2019b} to compute the planetary spectrum. Non-zero albedos could result in a shallow eclipse depth even in the absence of an atmosphere due to increased reflection of stellar radiation. Therefore, we include several simple blackbody airless scenarios for various albedos. We note that high albedos are deemed unlikely for airless planets as they would be subjected to severe space weathering, darkening their surface over the lifetime of the planets \citep{Domingue2014, Pieters2016, Mansfield2019}. The dayside brightness temperature $T_{\rm day}$ is computed based on the stellar effective temperature $T_{*}$, the stellar radius $R_{*}$, and the orbital distance $d$ using :
\begin{equation}\label{eq:Tday}
    T_{\rm day} = T_{*}\sqrt{\frac{R_{*}}{d}}(1-A_\mathrm{B})^{1/4}f^{1/4},
\end{equation}
with a redistribution factor $f=2/3$, which corresponds to the no heat redistribution limit, and varying the Bond albedo $A_\mathrm{B}$. For a completely dark surface without heat redistribution ($A_\mathrm{B} = 0.0$, $f=2/3$) the dayside temperature is T$_{\rm day}^{\rm max}=764~$K.

\section{System parameters} \label{sec:systemparameters}
The premise of secondary eclipse photometric observations to identify planetary atmospheres relies on a precise and accurate knowledge of the stellar and planetary system parameters. No single data set can provide constraints on all system parameters, but by combining transit, radial velocity, and photometric data with empirical constraints on the stellar mass and radius, a complete picture of the system emerges. Time series light curves from four TESS Sectors and RV measurements of LHS 1478 are already published \citep{Soto2021}, but no eclipses have been observed until now. Precise transit and eclipse times are particularly able to constrain orbital eccentricities through $e$cos$w$, while precise transit and eclipse durations can constrain $e$sin$w$ \citep{Mahajan2024}.

We perform a global fit of the LHS 1478 system with \texttt{ExofastV2} \citep{Eastman2013,Eastman2019} using all currently available TESS data (Sectors 18, 19, 25, 26, 52, 53, 59, 73, 79) at 120~s cadence to constrain the transit, the CARMENES and IRD data used by \citet{Soto2021} to constrain the radial velocity, photometric data to constrain the stellar SED, and empirical $M_\mathrm{K_s}$--$R_\mathrm{s}$ and $M_\mathrm{K_s}$--$M_\mathrm{s}$ relationships to constrain the stellar radius and mass, respectively \cite{Mann2015,Mann2019}. We also include the secondary eclipse measured in this work in the global fit. Due to the strong systematics observed in the second secondary eclipse visit, only the JWST/MIRI 15$~\upmu$m light curve from the first visit is used.

By fitting the planet and stellar parameters simultaneously, we achieve precise, up-to-date, and self-consistent results for the LHS 1478 system. Median and 68\% confidence interval values can be found in Table~\ref{tab:lhs1478}. We use an MCMC to explore the parameter space. We run the MCMC for 7500 steps with $n_\mathrm{thin} = 100$. We achieve a Gelman-Rubin statistic $\hat{r} < 1.01$ for all parameters except for the planetary inclination cos($i$), which is not well constrained by the precision of our data. Nevertheless, $\hat{r} = 1.0112$ for this parameter, which is close to our convergence criterion of $\hat{r} < 1.01$, and we therefore take this as meeting our convergence standards.

This joint fit on RV, transit and eclipse data allows us to refine the eccentricity and argument of periastron to $e = 0.038^{+0.160}_{-0.033}$ and $\omega = 86.2^{+4.5}_{-130.0}$ deg, however we caution that in order to reach convergence, we placed a prior of $\pm18$ min on the time of eclipse, referenced from the predicted eclipse time for a circular orbit. We do not have complete phase coverage of LHS 1478 at 15$\upmu$m, and the secondary eclipse is too shallow to be detected in the TESS data. The MCMC is therefore unable to explore the full parameter space to determine the maximum likelihood value for detecting the secondary eclipse; if the time of the secondary eclipse is allowed to wander outside of the JWST 15$\upmu$m data, there is no data to constrain the times of secondary eclipse. Because the eclipse depth is relatively shallow in the first visit of the JWST data, the MCMC does indeed wander outside of the available data if a prior is not applied. A such, the \texttt{ExofastV2} fit is not an independent line of evidence for favouring an eclipse over a flat line in the JWST 15$\upmu$m data; we discuss this further in Section~\ref{subsec:shlloweclipse}.
\section{Results}\label{sec:results}
Figure \ref{fig:atm_models} shows the eclipse depth $f_p/f_*$ inferred from the first observation in context with different atmospheric and bare rock models. We show 3 main types of atmospheres: N$_2$ dominated with the addition of $1-100\thinspace$ppm of CO$_2$ at $1\thinspace$bar (green group), mixed H$_2$O/CO$_2$ scenarios with different concentration combinations at $1\thinspace$bar (blue group), and finally pure CO$_2$ atmospheres at $0.1-10\thinspace$bar (pink group). The airless cases are modelled as blackbodies using Equation \eqref{eq:Tday} and varying the Bond albedo. The resulting dayside brightness temperature $T_{\rm day}$ is indicated in the legend for each scenario. In order to give a more accurate representation of the photometric measurement compared to the different models, we integrate each of them over the F1500W bandpass and show the resulting values in the right-hand panel, plotted over the final value for $f_p/f_*$ and its error bars (gray shaded region). 

If we take the eclipse depth inferred from the first visit as the planet’s dayside thermal emission, we can reject the null hypothesis (dark bare rock, no atmosphere) at about $3.3\upsigma$. We can reject a $A_{\rm B}=0.1$ albedo planet with no atmosphere at $2.7\upsigma$, and a $A_{\rm B}=0.2$ albedo planet with no atmosphere at only $2.1\upsigma$. Because of the nature of the light curve from the second observation, it is difficult to incorporate it into these statistics. This also suggests that additional observational follow-up is needed to confirm or reject our results. Almost all the atmospheric models considered fall within the error bars of the tentative eclipse depth over the full band pass (see right panel in Figure \ref{fig:atm_models}).

\begin{figure}
    \centering
    \includegraphics[width=\linewidth]{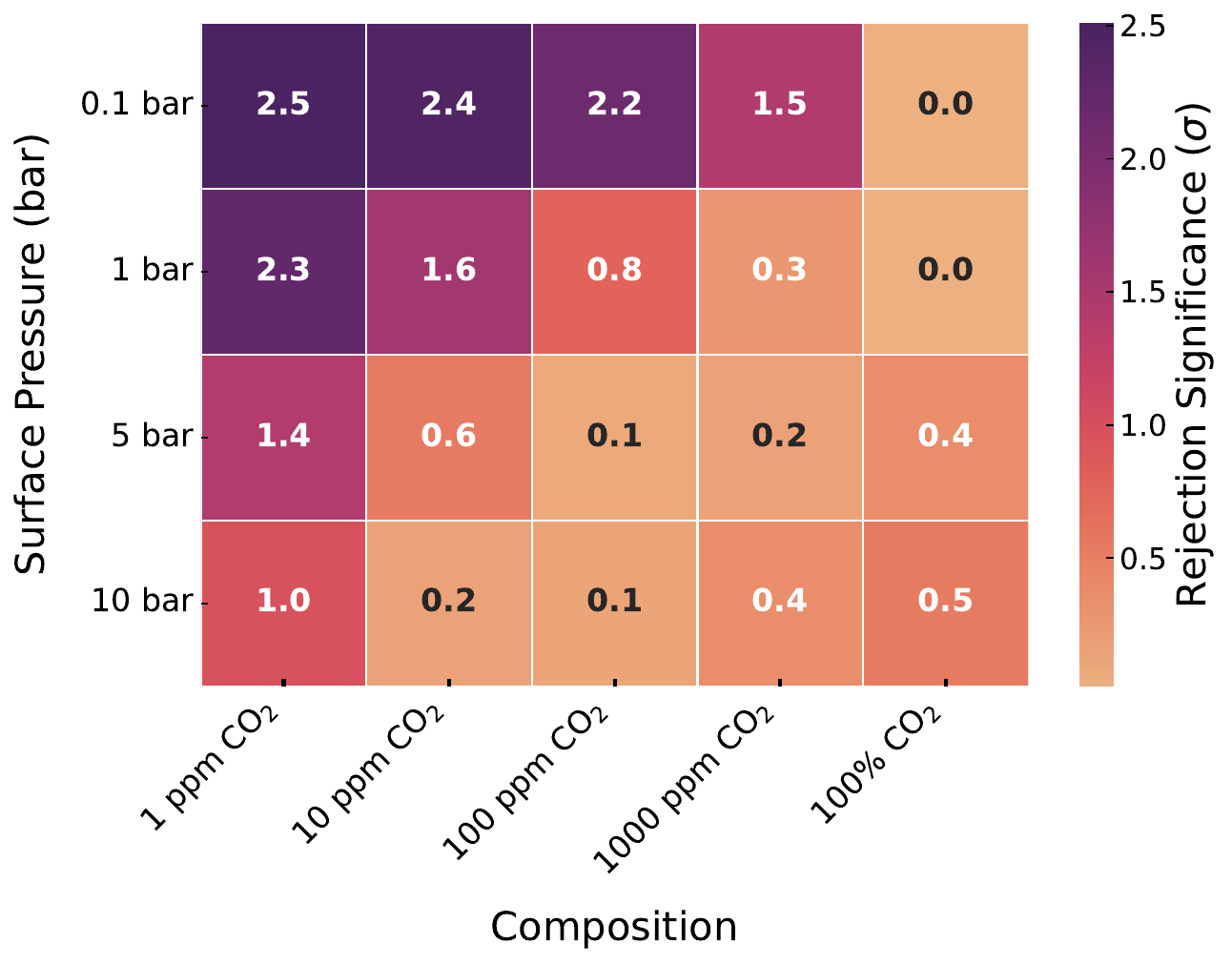}
    \caption{Grid plot showing the rejection significance (in $\sigma$) of different N$_2$/CO$_2$ atmospheric models based on the measured eclipse depth from the first observation. We vary CO$_2$ concentration ($1-1000~$ppm and a pure CO$_2$ atmosphere) on the x-axis and surface pressure ($1-10$ bar) on the y-axis. Higher values correspond to a higher rejection significance of the scenario considered.}
    \label{fig:N2CO2_sigmas}
\end{figure}

We focus on N$_2$-based atmospheres with CO$_2$ as a trace species in Figure \ref{fig:N2CO2_sigmas}. We find that atmospheres with CO$_2$ concentrations $<10\thinspace$ppm and surface pressures $<1\thinspace$bar are rejected at the $>2\upsigma$ level. Conversely, atmospheres with CO$_2$ concentrations $>10\thinspace$ppm and surface pressures $>5\thinspace$bar agree with the tentative eclipse depth at $<1\upsigma$. $10\thinspace$bar atmospheres need concentrations of $10~$ppm and above to reach the same level of agreement. This lower limit goes up to $100~$ppm for $1\thinspace$bar atmospheres, while very thin atmospheres require pure CO$_2$ composition to agree within $<1\upsigma$ with the observed data point. Pure CO$_2$ atmospheres agree at $<1\upsigma$ regardless of the pressure. 

All the pure CO$_2$ atmospheres considered, from $0.1$ to $10\thinspace$bar, are consistent with the tentative eclipse depth to within $1\upsigma$, the best fitting one being a $1\thinspace$bar pure CO$_2$ atmosphere. The mixed CO$_2$/H$_2$O models at $1\thinspace$bar also all agree with the data, showcasing the degeneracy the $15\upmu$m data point poses regarding more diverse atmospheric chemical compositions. Finally, against a neutral background gas like N$_2$, the addition of CO$_2$ introduces a sharp feature at $15\upmu$m. Models integrated over the whole band pass suggest that these agree less well with the photometric data point, falling outside of the $1\upsigma$ eclipse depth uncertainty, suggesting the need for higher concentrations of CO$_2$, or higher surface pressures, to match the candidate eclipse depth. For further airless scenarios, we considered high albedo surfaces that could reproduce the low eclipse depth. We find that albedos as high as $A_\mathrm{B}=0.4$ ($T_{day} = 672~$K) are needed to match the data point provided by the 15 $\upmu$m photometry at the $1\upsigma$ level. The data point is best fit by a surface Bond albedo of $A_\mathrm{B}=0.57$ ($T_{day} = 620~$K).

Lastly, we compute the $R = T_{b}/T_{day}^{max}$ factor described in \cite{Xue2024, Zhang2024}. To this end, we model the planetary emission with a null albedo blackbody ($A_{B}$) and no heat redistribution $(f=2/3)$. We then divide by the stellar spectrum, multiply by the planet-to-star radius ratio squared, and fit the resulting planetary spectrum to the observed data point at $15\upmu$m to obtain the brightness temperature $T_{b}$ using \texttt{emcee} \citep{emcee}. We obtain a best-fit brightness temperature of $T_{b} = 491 \pm 102~$K, yielding $R = 0.64 \pm 0.13$.

\section{Discussion}\label{sec:discussion}
\subsection{Data reduction}
One general take-away is that the JWST/MIRI Imaging observations seem to exhibit more systematics than expected.
The size of an error bar in the light curves is of $600~$ppm for both \texttt{Frida} light curves, and the RMS of the residuals is $714~$ppm for observation 1 and $705~$ppm for observation 2, giving scatter-to-error bar ratios $<1.2$. The different diagnostics (reduced $\chi^2$ metric, Allan plots) also suggest the white noise is well estimated, and not a problem in our observations. However, there seems to be time-correlated noise in addition to the well-known detector settling ramps and instrumental white noise.
In our case, the second observation was particularly affected, both by a strong initial ramp, and also by an unexplained, sinusoidal-looking signal occurring around the time of eclipse, with an amplitude of a couple hundreds of ppm and a timescale of about an hour. This signal, highlighted in red in Figure \ref{fig:lightcurves}, mimicked an eclipse about $35\thinspace$min later than predicted. Injection recovery tests showed that a signal similar to the first eclipse occurring at the expected time would be undetectable in the light curve due to this systematic. In fact, injected eclipse depths shallower than $300\thinspace$ppm occurring at the predicted time of eclipse could not be recovered due to the specific "up-and-down" pattern coinciding with the mid-eclipse and egress. This systematic is not correlated with any of the centroid positional information, nor with the guide star information which we inspected using \texttt{spelunker} \citep{Deal2024}. If it was not for that specific systematic, the eclipse signal might have been extracted from the light curve, as shown by the injection recovery tests where signals were injected in the light curve at a more "well-behaved" timing (see Table \ref{tab:injection_recoveries}). For this reason, the joint fits considering the two eclipses also effectively dilute the first observation's signal, shown by the results in \ref{subsec:jointfit}. 

The first observation is much more stable, and an eclipse was tentatively measured at the expected time and with the expected duration for a circular orbit. We find a candidate eclipse depth of $f_{p}/f_{*;f} = 138 \pm 53\thinspace$ppm with \texttt{Frida} and $f_{p}/f_{*;e} = 125 \pm 56\thinspace$ppm with \texttt{Eureka!}. The independent analysis using \texttt{transitspectroscopy} for the data reduction and \texttt{juliet} for the fitting returns a value of $f_p/f_{*;j} = 107 \pm 65\thinspace$ppm. All three values, and more broadly all values showcased in Table \ref{tab:data_reduction} are consistent within $1\upsigma$ of each other. The differences can partly be attributed to the optimal extraction method used in \texttt{Frida}, returning smaller RMS and MAD values for both light curves than the aperture photometry method (see Table \ref{tab:datared}) used in \texttt{Eureka!}. The \texttt{transitspectroscopy} reduction uses a version of optimal extraction on a 10-pixel aperture. This means including the secondary "petals" in addition to the central part of the PSF, as done in \texttt{Eureka!} (5-pixel aperture). The secondary diffraction patterns are difficult to separate from the background in this area and including them, even if weighted down, could introduce more background flux, which is prevented by \texttt{Frida}'s "z-cut" method. Finally, the \texttt{Eureka!} reduction evaluates the background in an annulus around the PSF rather than on the full detector, and the \texttt{Frida} "z-cut" and optimal weights method allows to disregard background subtraction. This might also introduce differences in the light curve depending on how homogeneous the background is across the detector and close to the PSF. 
Overall, this effort highlights the need for multiple data reductions to cross-validate results, as well as careful monitoring of the different aspects of the light curve extraction, especially because of the complex shape of the JWST/MIRI PSF.

We utilise the value obtained through the \texttt{Frida} data reduction and fitting of the first visit with $n_{clip}=100$ integrations removed at the beginning of the light curve and a simple linear model. The optimal extraction with the "z-cut" method as implemented in \texttt{Frida} yielded the best results in terms of RMS and MAD values (see Table \ref{tab:datared}) compared to the other data reductions. Interestingly, the second visit gets even lower RMS and MAD values from Frida and Eureka, demonstrating the limitations of these metrics in the presence of correlated noise. The different systematic models and detector settling removal combinations considered in Table \ref{tab:data_reduction} lead to very similar BIC and $\chi^2_{\rm reduced}$ values and agreeing eclipse depths. We thus opt for the simplest model and a clipping which removes all of the initial ramp while keeping as much baseline as possible.

\subsection{Shallow eclipse or missed eclipse?}\label{subsec:shlloweclipse}
Since we detect the eclipse at $2.6\upsigma$ in the first observation and are unable to detect the eclipse in the second observation, we explore whether we could have missed the eclipse entirely due to changes in orbital eccentricity of LHS~1478~b. For a short-period planet like LHS~1478~b, tidal circularisation occurs very rapidly on timescales of thousands to millions of years depending on the initial eccentricity \citep{Raymond2008}. It is difficult to determine the ages of M dwarfs, but their rotation periods can serve as an age proxy since they spin down over time \citep[e.g.,][]{Engle2023}. Attempts to measure the stellar rotation period of LHS~1478 have yielded non-detections, but the stellar activity indicators suggest a relatively quiet star \citep{Soto2021, Newton2016}. While we cannot estimate the age based on the non-detected stellar rotation period, the data suggest it is longer than two TESS sectors, given the target is observed consecutively in sectors 18 and 19, 25 and 26, and 52 and 53. Based on work by \cite{Medina2022b} and \cite{Engle2023}, this would imply that the age of the LHS 1478 star and system is about $5.6\pm 2.7~$Gyr, which is much longer than the circularisation timescale. It is therefore highly likely that the orbit of LHS~1478~b has been circularised, unless eccentricity is being induced by tidal effects or massive outer companions. The RVs for this system do not allow for such a companion \citep{Soto2021}.

To confirm that the non-detection in the second observation is not due to tidal effects modifying the orbit between the two visits, and hence eclipse times, we performed simulations of the system with VPLanet \citep{Barnes2020}. We considered two equilibrium tide models (often called the constant-phase-lag and constant-time-lag models \citep{Greenberg2009}), set the tidal dissipation to be equal to modern Earth's \citep{Lambeck1977,Williams1978}, and set the initial eccentricity to 0.5. All other parameters were fixed to the best-fit values from \cite{Soto2021}. We find that over a 1-month timescale, the eccentricity can only change by about 1 ppm for these extreme conditions, confirming that tides are not affecting the eclipse ephemerides between the two visits.
Note that this assessment assumes LHS~1478~b has no companions, which could significantly alter the circularisation timescale \citep[see e.g.,][]{Livesey2024,Barnes2024}. \\

\subsection{Origin of the systematic in the second observation}
We found no correlation between the systematic in the second observation and the shift of the centroid of the star in $x$ and $y$ position, nor with the guide star information. Further investigating an instrumental origin, we Fourier-transformed each individual pixel light curve of both observations into frequency space. No significant time-correlated signals were identified, and a K-means clustering algorithm \citep[\texttt{sklearn};][]{Pedregosa2011} on the frequency-space time-series also did not reveal any detector-wide patterns. While the sinusoidal systematic observed in the second visit could still be instrumental, it is possible that it has an astrophysical origin. 

The characteristic shape resembles similar systematics observed in infrared secondary eclipse light curves of 55~Cancri~e, linked to the planet's volcanic outgassing and surface activity \citep{Demory2016, Patel2024}. TESS observations for 55~Cnc~e also showed variability in the phase curve and eclipse depth measurements, but not the primary transits \citep{MeierValdes2023}. This raises the question of whether similar mechanisms, such as atmospheric variability, could affect the observations of LHS~1478~b. 

Another astrophysical source could be stellar activity, although TESS photometric and CARMENES spectroscopic observations suggest LHS~1478 to be a fairly inactive star \citep{Soto2021}. Emerging JWST observations suggest that stellar flaring activity may be detectable even at $15~\upmu$m, challenging previous assumptions about its negligible impact in this regime. If such flares influence secondary eclipse measurements, their potential effects remain uncertain, as their signatures in the mid-infrared are not well characterised.

Regardless, our ability to investigate whether the observed variability is periodic or linked to transient phenomena is limited with only two eclipse observations which lack long temporal resolution. Observing simultaneous eclipses in multiple infrared wavelengths, as was done by \citet{Patel2024}, could be helpful in understanding the origin of the systematic visible in the second observation. With the current data, we cannot definitively attribute the observed systematic to either an instrumental or astrophysical origin.

\subsection{Model interpretation}
Based on the first visit, we infer a tentative eclipse depth of $f_{p}/f_{*;f} = 138 \pm 53\thinspace$ppm, a $2.6\upsigma$ evidence of an eclipse. If the tentative eclipse depth is representative of the planet's emission, it would be lower than the predicted bare rock, $A_{\rm B}=0.1$ albedo no atmosphere model at $2.7\upsigma$, and lower than the bare rock, $A_{\rm B}=0.2$ albedo no atmosphere model at $2.1\upsigma$. However, given that we only detect the eclipse in the first visit at $2.6\upsigma$, and the non-detection in the second visit, we cannot reject the no-atmosphere scenario based on the current data. 
This highlights the importance of observing  these lower signal targets multiple times to ensure the repeatability and consistency of the detections. This is particularly true in the early JWST era, as the community still needs time to address and understand the systematics affecting these new instruments.

The high albedos ($A_{B}>0.4$) necessary to match the photometric data at $15\thinspace\upmu$m with bare rock models are not expected from theory or observations. This is mostly due to space weathering \citep{Pieters2016}, which darkens the surface over time. Overall, albedos $>0.2$ are unlikely for airless rocky planets \citep{Domingue2014}. However, it is worth noting that space weathering and resurfacing processes are a topic of ongoing research in the field. Follow-up studies considering the diverse range of rock surface albedos are necessary to examine possible surface compositions of this planet and place further constraints on the possibility of an atmosphere.

\section{Conclusions}\label{sec:conclusions}
LHS-1478~b was observed during secondary eclipse on two occasions with JWST/MIRI F1500W. The first visit reveals a possible eclipse at the expected time for a circular orbit (within the propagated error bars) with a depth of $f_{p}/f_{*} = 138 \pm 53\thinspace$ppm, preferred over a flat line model at $2.8\upsigma$. The non-informative priors on the time of secondary eclipse and eclipse depth, as well as the circularisation timescale arguments for this system, further support a detection. The injection recovery tests, which show that the same signal from the first observation would not be recoverable in the second, lead us to believe that the non-detection in the second observation is likely due to instrumental or astrophysical systematics rather than a missed eclipse.

We find that an eclipse depth of $f_{p}/f_{*} = 138 \pm 53\thinspace$ppm is shallower than most airless, low-albedo emission scenarios for this planet. If this represents the true eclipse depth, we can reject the null hypothesis of the dark (zero albedo) bare rock model with a confidence level of $3.3\upsigma$. This number decreases to $2.1\upsigma$ for an airless rock with a surface albedo of $A_{\rm B}=0.2$. A caveat is that we used simple models of spectrally neutral surfaces to represent the airless scenarios. Follow-up studies considering a range of rocky surface models can constrain possible surface compositions for this planet.

Most atmospheric models considered, covering different types of CO$_2$ atmospheres (pure, mixed with H$_2$O, and as a trace species in an N$_2$-based atmosphere), are consistent within the error bars of the secondary eclipse depth measurement from the first visit, except for cases with very little CO$_2$ and surface pressures of $1$ bar or lower.

However, we emphasise that the two observations from our program do not yield consistent results. Therefore, additional observations are necessary to confirm the measurement from the first observation, especially because the interpretation of whether the planet has an atmosphere or not relies on a single photometric data point from a single observation. If the second observation was plagued by systematics, we cannot be certain that the first observation is unaffected by systematics, or that the eclipse was definitively detected. Consequently, additional observations are necessary to ensure the reproducibility of the results. Additionally, broader spectral coverage would allow us to disentangle different atmospheric or surface scenarios and shed more light on the chemistry of this planet.

This dataset illustrates the need for a larger community effort to improve our understanding of the systematics affecting MIRI, and JWST detectors in general. Our results underscore the challenge of using eclipse photometry to detect rocky planet atmospheres as opposed to bare rocks. Shallow eclipses are more difficult to detect with statistical significance, especially in the presence of time-correlated noise. Precise knowledge of ephemerides and eccentricities is therefore crucial to avoid interpreting a missed eclipse. This is maybe especially important given the large amount of time dedicated to this technique with the upcoming DDT Rocky Worlds program. 

\begin{acknowledgements}
Based on observations with the NASA/ESA/CSA James Webb Space Telescope
obtained at the Space Telescope Science Institute, which is operated by the Association of Universities for Research in Astronomy, Incorporated, under NASA contract NAS5-03127.
PCA and HDL acknowledges support from the Carlsberg Foundation, grant CF22-1254. 
JMM acknowledges support from the Horizon Europe Guarantee, grant EP/Z00330X/1.
ADR acknowledges support from the Carlsberg Foundation, grant CF22-1548.
NHA acknowledges support by the National Science Foundation Graduate Research Fellowship under Grant No. DGE1746891.
EMV aknowledges support from the Centre for Space and Habitability (CSH). This work has been carried out within the framework of the National Centre of Competence in Research PlanetS supported by the Swiss National Science Foundation under grant 51NF40\_205606. EMV acknowledges the financial support of the SNSF.
B.-O. D. acknowledges support from the Swiss State Secretariat for Education, Research and Innovation (SERI) under contract number MB22.00046.
CEF acknowledges support from the European Research Council (ERC) under the European Union's Horizon 2020 research and innovation program under grant agreement no. 805445.
NPG gratefully acknowledges support from Science Foundation Ireland and the Royal Society through a University Research Fellowship (URF\textbackslash R\textbackslash 201032).
HJH and BP acknowledges support from eSSENCE (grant number eSSENCE@LU 9:3), The Crafoord foundation and the Royal Physiographic Society of Lund, through The Fund of the Walter Gyllenberg Foundation.
We acknowledge the input from the following individuals to the GO 3730 proposal: Adam Burgasser, Can Akin, Andrea Guzm\a'an Mesa, Nicholas Borsato, Meng Tian, Mette Baungaard.
\end{acknowledgements}

\bibliographystyle{aa}  
\bibliography{aanda}  

\begin{appendix}
    \section{Diagnostics of the light curves}\label{app:1}
\begin{figure}[h]
        \centering
        \includegraphics[width=\linewidth]{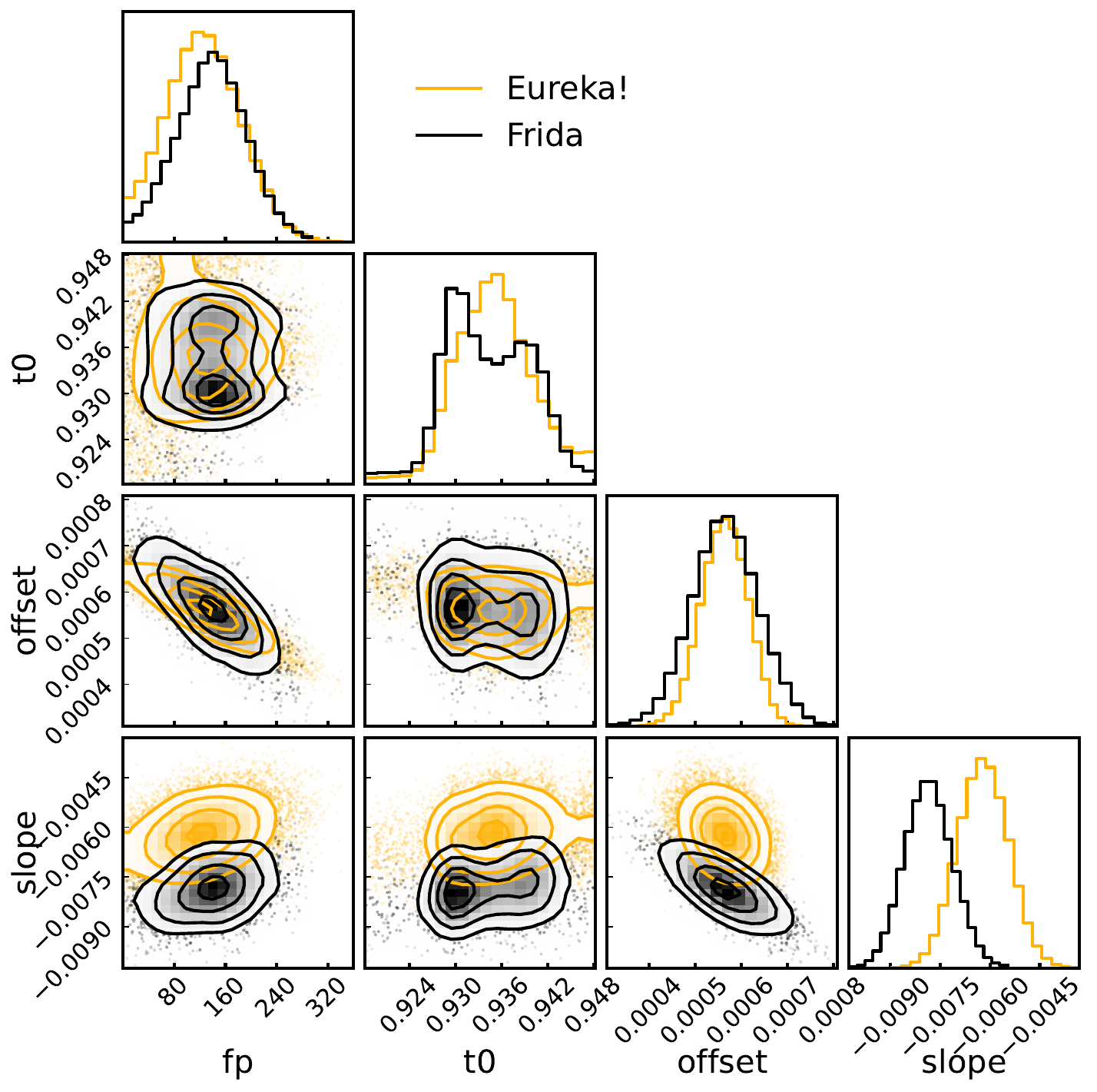}
        \caption{Corner plot for the \texttt{Frida} (black) and \texttt{Eureka!} (orange) light curve fits of the first observation.}
        \label{fig:cornerplot}
    \end{figure}

    Here we provide some additional diagnostics to the light curves as obtained from the \texttt{Frida} and \texttt{Eureka!} data reductions and light curve fits.
    In Figure \ref{fig:cornerplot}, we present the superimposed \texttt{Frida} and \texttt{Eureka!} corner plot of the light curve model highlighted in Table \ref{tab:data_reduction}, that is a simple linear slope with the initial $n_{clip}=100$ integrations removed. The offset from \texttt{Eureka!} is adjusted with a constant to match the \texttt{Frida} convention for visual clarity. 
    
\begin{figure}[h]
        \centering
        \includegraphics[width=\linewidth]{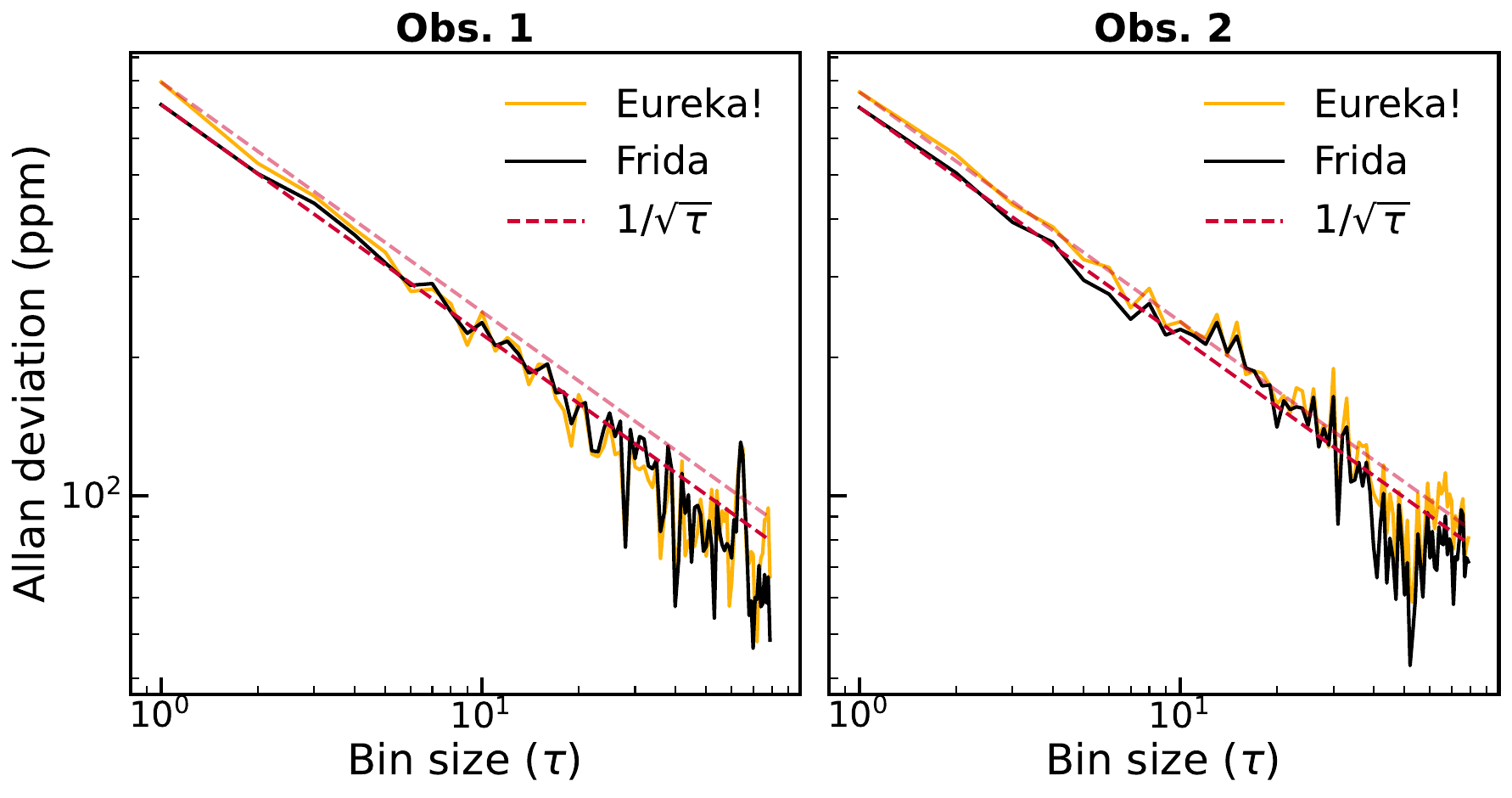}
        \caption{Allan deviation plot for the \texttt{Frida} and \texttt{Eureka!} reductions and fits for both observations.}
        \label{fig:allandeviation}
    \end{figure}

    We also show the Allan deviation plots for both light curves in Figure \ref{fig:allandeviation}. For the second observation, we discard the first $n_{clip}=150$ to remove the detector settling slope, and use an exponential with a linear slope (\texttt{Frida}), respectively a simple linear slope (\texttt{Eureka!}), to fit the light curve. This plot shows the RMS of the residuals as a function of the bin size $\tau$. In the absence of correlated noise, the orange and black lines should follow the red dashed line, representing the white noise decreasing as $1/\sqrt{\tau}$.


    \section{Edge cases for the heat redistribution factor $f$}\label{app:2}
    In the models shown in Figure \ref{fig:atm_models}, $f$ is obtained based on \cite{Koll2022} through an estimation of the long-wave optical depth of the atmosphere, which in turn also depends on atmospheric properties. This requires several re-runs to converge. In the interest of completeness regarding the estimation of that parameter, we recompute several of the highlighted models, artificially fixing $f$ to its "edge case" values, that is $0.25$ for total heat redistribution and $2/3$ for no heat redistribution.
    \begin{figure}[h]
        \centering
        \includegraphics[width=\linewidth]{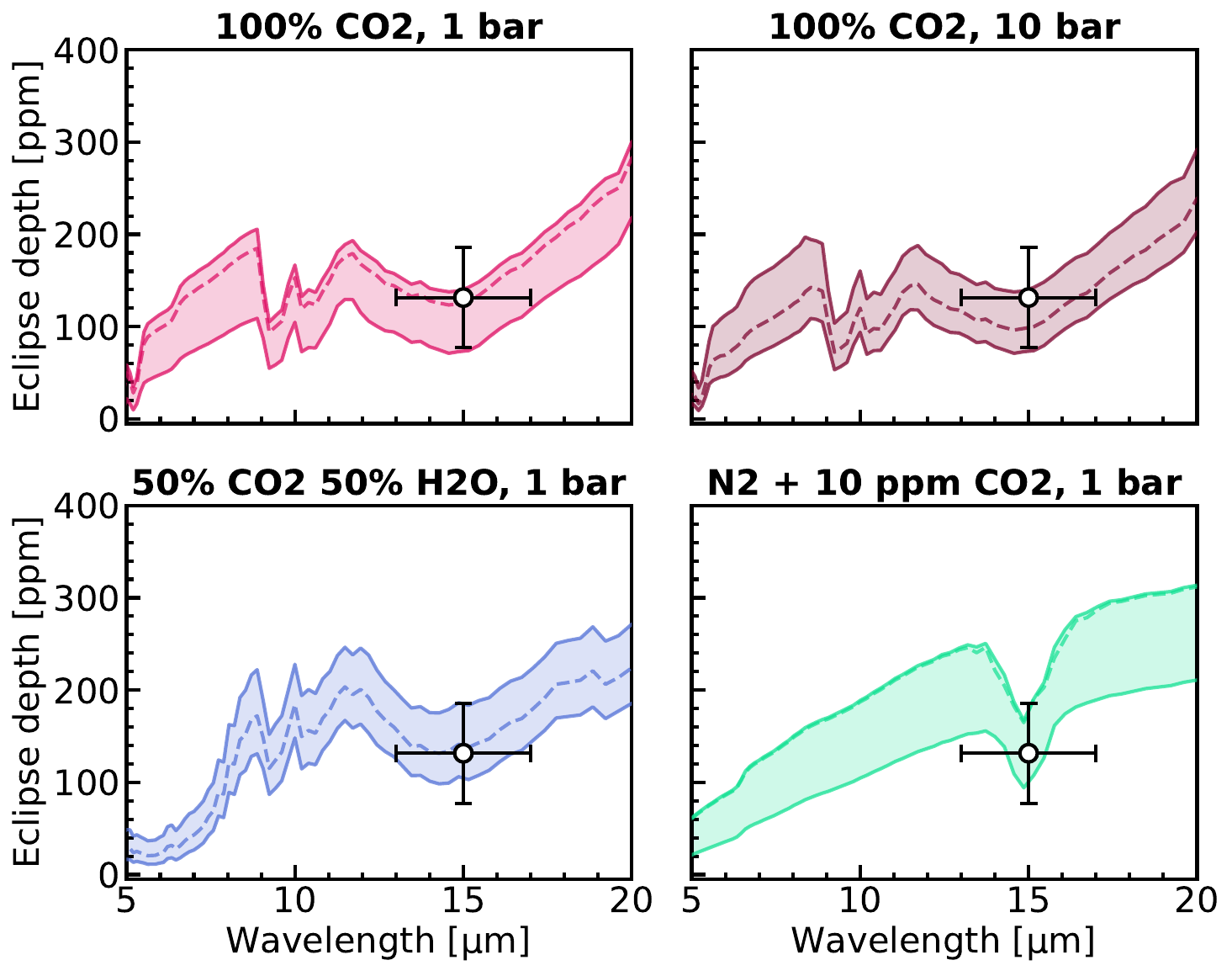}
        \caption{Impact on the modelled atmospheric spectra of extreme values of the redistribution factor $f$. The lower lines represent $f=0.25$ while the upper lines represent $f=2/3$. The dashed line is the corresponding model shown on Figure \ref{fig:atm_models}.}
        \label{fig:edgecasesf}
    \end{figure}

    \section{Global parameter fit for LHS 1478}\label{app:3}
    In Table~\ref{tab:lhs1478} provide the best-fit values and 68\% confidence intervals for parameters in a global fit of the LHS 1478 star-planet system using \texttt{ExovastV2}, commit \texttt{6ba004d}.
    \newpage
    \onecolumn
    \setstretch{1.15}
    \setlength{\tabcolsep}{1pt}
    \begin{longtable}{l@{\hskip 4pt}l@{\hskip 4pt}c@{\hskip 4pt}c@{\hskip 4pt}c@{\hskip 4pt}c@{\hskip 4pt}}
    \caption{\texttt{EXOFASTv2} median values and 68\% confidence interval for LHS 1478}\label{tab:lhs1478} \\
    \hline\hline
    Parameter & Description & \multicolumn{4}{c}{Values} \\
    \hline
    \vspace{5mm}
    \endfirsthead 
    \caption{continued.}\\ 
    \hline\hline 
    Parameter & Description & \multicolumn{4}{c}{Values} \\
    \hline
    \endhead
    \hline
    \endfoot
    Stellar Parameters:& & LHS 1478 & & \smallskip\\
    ~~~~$M_*$ & Mass ($M_{\odot}$)\dotfill &$0.2301\pm0.0054$\\
    ~~~~$R_*$\dotfill &Radius ($R_{\odot}$)\dotfill &$0.2462^{+0.0080}_{-0.0079}$\\
    ~~~~$L_*$\dotfill &Luminosity ($L_{\odot}$)\dotfill &$0.00742^{+0.00053}_{-0.00032}$\\
    ~~~~$F_{Bol}$\dotfill &Bolometric Flux (cgs)\dotfill &$7.16^{+0.51}_{-0.31} \times 10^{-10}$\\
    ~~~~$\rho_*$\dotfill &Density (cgs)\dotfill &$21.7^{+2.3}_{-2.0}$\\
    ~~~~$\log{g}$\dotfill &Surface gravity (cgs)\dotfill &$5.017^{+0.030}_{-0.029}$\\
    ~~~~$T_{\rm eff}$\dotfill &Effective temperature (K)\dotfill &$3415^{+82}_{-63}$\\
    ~~~~$[{\rm Fe/H}]$\dotfill &Metallicity (dex)\dotfill &$-0.38^{+0.20}_{-0.19}$\\
    ~~~~$K_S$\dotfill &Absolute Ks-band mag (mag)\dotfill &$7.462\pm0.022$\\
    ~~~~$k_S$\dotfill &Apparent Ks-band mag (mag)\dotfill &$8.764\pm0.022$\\
    ~~~~$A_V$\dotfill &V-band extinction (mag)\dotfill &$0.135^{+0.21}_{-0.100}$\\
    ~~~~$\varpi$\dotfill &Parallax (mas)\dotfill &$54.909\pm0.020$\\
    ~~~~$d$\dotfill &Distance (pc)\dotfill &$18.2120^{+0.0067}_{-0.0068}$\\
    \smallskip\\\multicolumn{2}{l}{Planetary Parameters:}&b\smallskip\\
    ~~~~$P$\dotfill &Period (days)\dotfill &$1.94953941\pm0.00000050$\\
    ~~~~$R_P$\dotfill &Radius (\re)\dotfill &$1.174\pm0.055$\\
    ~~~~$M_P$\dotfill &Mass (\me)\dotfill &$2.27\pm0.45$\\
    ~~~~$T_C$\dotfill &Observed Time of conjunction$^{2}$ (\bjdtdb)\dotfill &$2458786.75416^{+0.00024}_{-0.00023}$\\
    ~~~~$T_0$\dotfill &Obs time of min proj sep$^{4,6,7}$ (\bjdtdb)\dotfill &$2459492.48743\pm0.00015$\\
    ~~~~$a$\dotfill &Semi-major axis (AU)\dotfill &$0.01872\pm0.00015$\\
    ~~~~$i$\dotfill &Inclination (Degrees)\dotfill &$87.69^{+0.41}_{-0.22}$\\
    ~~~~$e$\dotfill &Eccentricity \dotfill &$0.038^{+0.16}_{-0.033}$\\
    ~~~~$\omega_*$\dotfill &Arg of periastron (Degrees)\dotfill &$86.2^{+4.5}_{-130}$\\
    ~~~~$\dot{\omega}_{\rm GR}$\dotfill &Computed GR precession ($^\circ$/century)\dotfill &$2.478^{+0.083}_{-0.048}$\\
    ~~~~$T_{\rm eq}$\dotfill &Equilibrium temp$^{8}$ (K)\dotfill &$597.3^{+11}_{-7.0}$\\
    ~~~~$\tau_{\rm circ}$\dotfill &Tidal circ timescale (Gyr)\dotfill &$9.9^{+3.7}_{-3.1}$\\
    ~~~~$K$\dotfill &RV semi-amplitude (m/s)\dotfill &$3.12\pm0.62$\\
    ~~~~$R_P/R_*$\dotfill &Radius of planet in stellar radii \dotfill &$0.0439^{+0.0010}_{-0.0015}$\\
    ~~~~$a/R_*$\dotfill &Semi-major axis in stellar radii \dotfill &$16.35^{+0.55}_{-0.52}$\\
    ~~~~$\delta$\dotfill &$\left(R_P/R_*\right)^2$ \dotfill &$0.001927^{+0.000092}_{-0.00013}$\\
    ~~~~$\tau$\dotfill &In/egress transit duration (days)\dotfill &$0.00211^{+0.00032}_{-0.00064}$\\
    ~~~~$T_{14}$\dotfill &Total transit duration (days)\dotfill &$0.02972^{+0.00051}_{-0.00052}$\\
    ~~~~$T_{FWHM}$\dotfill &FWHM transit duration (days)\dotfill &$0.02768^{+0.00048}_{-0.00044}$\\
    ~~~~$b$\dotfill &Transit impact parameter \dotfill &$0.647^{+0.054}_{-0.20}$\\
    ~~~~$b_S$\dotfill &Eclipse impact parameter \dotfill &$0.683^{+0.047}_{-0.064}$\\
    ~~~~$\tau_S$\dotfill &In/egress eclipse duration (days)\dotfill &$0.00240^{+0.00036}_{-0.00024}$\\
    ~~~~$T_{S,14}$\dotfill &Total eclipse duration (days)\dotfill &$0.03040^{+0.0067}_{-0.00078}$\\
    ~~~~$T_{S,FWHM}$\dotfill &FWHM eclipse duration (days)\dotfill &$0.02800^{+0.0064}_{-0.00077}$\\
    ~~~~$\rho_P$\dotfill &Density (cgs)\dotfill &$7.7^{+2.0}_{-1.7}$\\
    ~~~~$logg_P$\dotfill &Surface gravity (cgs)\dotfill &$3.206^{+0.090}_{-0.10}$\\
    ~~~~$\Theta$\dotfill &Safronov Number \dotfill &$0.0111^{+0.0023}_{-0.0022}$\\
    ~~~~$\fave$\dotfill &Incident Flux (\fluxcgs)\dotfill &$0.0285^{+0.0021}_{-0.0015}$\\
    ~~~~$T_S$\dotfill &Observed Time of eclipse$^{2}$ (\bjdtdb)\dotfill &$2458785.7832^{+0.0072}_{-0.0047}$\\
    ~~~~$T_{E,0}$\dotfill &Obs time of sec min proj sep$^{4,6,7}$ (\bjdtdb)\dotfill &$2460483.8320^{+0.0072}_{-2.0}$\\
    ~~~~$T_P$\dotfill &Time of Periastron (\tjdtdb)\dotfill &$2458786.738^{+0.019}_{-0.71}$\\
    ~~~~$T_A$\dotfill &Time of asc node (\tjdtdb)\dotfill &$2458786.286^{+0.10}_{-0.021}$\\
    ~~~~$T_D$\dotfill &Time of desc node (\tjdtdb)\dotfill &$2458787.227^{+0.019}_{-0.10}$\\
    ~~~~$V_c/V_e$\dotfill &Scaled velocity \dotfill &$0.973^{+0.031}_{-0.15}$\\
    ~~~~$e\cos{\omega_*}$\dotfill & \dotfill &$0.0032^{+0.0057}_{-0.0037}$\\
    ~~~~$e\sin{\omega_*}$\dotfill & \dotfill &$0.027^{+0.17}_{-0.031}$\\
    ~~~~$M_P\sin i$\dotfill &Minimum mass (\me)\dotfill &$2.27\pm0.45$\\
    ~~~~$M_P/M_*$\dotfill &Mass ratio \dotfill &$2.96\pm0.58 \times 10^{-5}$\\
    \smallskip\\\multicolumn{2}{l}{Wavelength Parameters:}&14.98$\upmu$m&TESS\smallskip\\
    ~~~~$u_{1}$\dotfill &Linear limb-darkening coeff \dotfill &--&$0.35^{+0.19}_{-0.21}$\\
    ~~~~$u_{2}$\dotfill &Quadratic limb-darkening coeff \dotfill &--&$0.28^{+0.28}_{-0.20}$\\
    ~~~~$A_T$\dotfill &Thermal emission from the planet (ppm)\dotfill &$135^{+67}_{-68}$&--\\
    ~~~~$\delta_{S}$\dotfill &Measured eclipse depth (ppm)\dotfill &$135^{+67}_{-68}$&--\\
    \smallskip\\\multicolumn{2}{l}{Telescope Parameters:}&CARMENES& IRD\smallskip\\
    ~~~~$\gamma_{\rm rel}$\dotfill &Relative RV Offset (m/s)\dotfill &$0.31^{+0.45}_{-0.43}$&$-2.0^{+1.9}_{-1.8}$\\
    ~~~~$\sigma_J$\dotfill &RV Jitter (m/s)\dotfill &$1.99^{+0.52}_{-0.51}$&$4.7^{+2.2}_{-1.8}$\\
    ~~~~$\sigma_J^2$\dotfill &RV Jitter Variance \dotfill &$4.0^{+2.4}_{-1.8}$&$21^{+26}_{-14}$\\
    \smallskip\\\multicolumn{2}{l}{Transit Parameters:}&JWST UT HS14-78-b. (14.98$\upmu$m)& TESS UT 2019-11-03 (TESS) \smallskip\\
    ~~~~$\sigma^{2}$\dotfill &Added Variance \dotfill & $1.50^{+0.30}_{-0.28} \times 10^{-7}$&$-8.245\pm0.051 \times 10^{-6}$ \\ 
    ~~~~$F_0$\dotfill &Baseline flux \dotfill &$0.999772^{+0.000052}_{-0.000053}$&$1.0001\pm0.0030$ \\
    ~~~~$C_{0}$\dotfill &Additive detrending coeff \dotfill  &$-0.000323^{+0.000057}_{-0.000054}$&--  \\
    
    & & TESS UT 2019-11-28 (TESS) & TESS UT 2020-05-14 (TESS)&\\
    ~~~~$\sigma^{2}$\dotfill &Added Variance \dotfill &$-1.974\pm0.048 \times 10^{-6}$&$-1.182\pm0.047 \times 10^{-6}$ \\
    ~~~~$F_0$\dotfill &Baseline flux \dotfill &$1.0000\pm0.0031$&$1.0001\pm0.0024$\\
    
    & & TESS UT 2020-06-10 (TESS) & TESS UT 2022-05-19 (TESS)\\
    ~~~~$\sigma^{2}$\dotfill &Added Variance \dotfill &$-5.252^{+0.051}_{-0.050} \times 10^{-6}$&$2.45\pm0.43 \times 10^{-7}$ \\
    ~~~~$F_0$\dotfill &Baseline flux \dotfill &$1.0000\pm0.0031$&$1.0000\pm0.0029$\\
    
    & &TESS UT 2022-06-13 (TESS) & TESS UT 2022-05-19 (TESS)\\
    ~~~~$\sigma^{2}$\dotfill &Added Variance \dotfill &$-2.06^{+0.44}_{-0.43} \times 10^{-7}$&$-2.14^{+0.45}_{-0.44} \times 10^{-7}$ \\
    ~~~~$F_0$\dotfill &Baseline flux \dotfill &$0.9999^{+0.0031}_{-0.0030}$&$1.0000\pm0.0031$ \\
    
    & &TESS UT 2022-06-13 (TESS) & TESS UT 2022-11-28 (TESS)\\
    ~~~~$\sigma^{2}$\dotfill &Added Variance \dotfill & $-2.14\pm0.44 \times 10^{-7}$ & $-2.06^{+0.44}_{-0.43} \times 10^{-7}$\\
    ~~~~$F_0$\dotfill &Baseline flux \dotfill &$1.0000\pm0.0031$ & $0.9999^{+0.0031}_{-0.0030}$\\
    
    & &TESS UT 2023-12-07 (TESS) & TESS UT 2024-05-22 (TESS)  \\
    ~~~~$\sigma^{2}$\dotfill &Added Variance \dotfill &$-1.50^{+0.49}_{-0.48} \times 10^{-7}$&$-0.5\pm4.3 \times 10^{-8}$ \\
    ~~~~$F_0$\dotfill &Baseline flux \dotfill&$1.0000^{+0.0032}_{-0.0031}$&$0.9997^{+0.0033}_{-0.0030}$ \\
    \end{longtable}
    \tablefoot{}\\{See Table 3 in \citet{Eastman2019} for a detailed description of all parameters}
    \tablefoottext{1}{This value ignores the systematic error and is for reference only}
    \tablefoottext{2}{Time of conjunction is commonly reported as the ``transit time''}
    \tablefoottext{3}{\tjdtdb is the target's barycentric frame and corrects for light travel time}
    \tablefoottext{4}{Time of minimum projected separation is a more correct ``transit time''}
    \tablefoottext{5}{Use this to model TTVs, e}
    \tablefoottext{6}{At the epoch that minimises the covariance between $T_C$ and Period}
    \tablefoottext{7}{Use this to predict future transit times}
    \tablefoottext{8}{Assumes no albedo and perfect redistribution}
    
\end{appendix}

\end{document}